\documentclass[aps,reprint,superscriptaddress]{revtex4-1}

\usepackage{graphicx}
\usepackage{dcolumn}
\usepackage{bm}
\usepackage{siunitx}
\usepackage{hyperref}
\usepackage{color}
\usepackage[dvipsnames]{xcolor}
\usepackage{float}
\usepackage{amsmath}
\usepackage{physics}
\usepackage{dirtytalk}

\hypersetup{
colorlinks=true,
urlcolor=blue,
citecolor=blue,
linkcolor=black
}

\begin{document}
\title
{
Tunable mechanically-induced hysteresis in suspended Josephson junctions
}

\author{Jamie Le Signe}
\altaffiliation[These authors contributed equally to this work.]
{}
\affiliation
{
School of Physics and Astronomy, University of Exeter, EX4 4QL, Exeter, United Kingdom
}

\author{Thomas McDermott}
\altaffiliation[These authors contributed equally to this work.]{}
\affiliation
{
School of Physics and Astronomy, University of Exeter, EX4 4QL, Exeter, United Kingdom
}
\affiliation
{
Faculty of Physics, University of Warsaw, 02-093, Warsaw, Poland
}

\author{Eros Mariani}
\email[Corresponding author: ]{E.Mariani@exeter.ac.uk}

\affiliation
{
School of Physics and Astronomy, University of Exeter, EX4 4QL, Exeter, United Kingdom
}
\date{\today}

\begin{abstract}

The coupling of superconducting systems to mechanical resonators is an emerging field, with wide reaching implications including high precision sensing and metrology. 
Experimental signatures of this coupling have so far been small, seldom and often reliant on high frequency AC electronics.
To overcome this limitation, in this work we consider a mechanical resonator suspended between two superconducting contacts to form a suspended Josephson junction in which the electronic normal- and super-currents can be coupled to mechanical motion via the Lorentz force due to an external magnetic field.
We show both analytically and numerically that this electro-mechanical coupling produces unprecedented mechanically-induced hysteresis loops in the junction's DC I-V characteristic. 
Firstly, we unveil how this new hysteresis may be exploited to access a huge mechanically-induced Shapiro-like voltage plateau, extending over a current range comparable with the junction's critical current.
We then investigate a sudden mechanically-induced retrapping that occurs at strong coupling. Our analytical treatment provides a clear explanation for the effects above and allows us to derive simple relationships between the features in the DC I-V characteristic and the resonance frequency and quality factor of the mechanical resonator.
We stress that our setup requires only DC current bias and voltage measurements, allowing the activation and detection of high-frequency mechanical oscillations in state of the art devices and without the need of any AC equipment.
\end{abstract}

\maketitle



\section{Introduction}
\label{sec:Introduction}

The recent progress of nanofabrication techniques has led to a rapid expansion of the nano-electromechanical systems (NEMS) industry, spurring the creation of nano-scale actuators \cite{dong}, motors \cite{kim} and switches \cite{jang}. 
Various groups have used carbon allotropes such as diamond \cite{gaidarzhy}, carbon nanotubes \cite{laird,moser} and graphene \cite{bunch,chen} to create mechanical resonators with resonance frequencies in the GHz range \cite{laird} and quality factors greater than $10^6$ \cite{moser}. These sharp, high frequency  resonators find applications in attogram mass measurement \cite{ilic,li}, single molecule detection \cite{naik} and high density data storage \cite{despont}.
Besides the plethora of promising applications, NEMS also provide a playground to explore the fundamental properties of mesoscopic devices \cite{bachtold}, including the quantization of heat transfer \cite{schwab} and the cooling of macroscopic oscillators to their quantum mechanical ground state \cite{sonne,sonne2}.
An area yet to be thoroughly explored is the interplay between electronic and mechanical degrees of freedom in superconducting systems \cite{haque,kretinin,marchenkov,schneider,etakiSSO,Khosla}. Experimental signatures of this interplay have so far been small, seldom and often reliant on high frequency AC electronics. 

\begin{figure}
\begin{centering}
    \includegraphics[width=80mm]{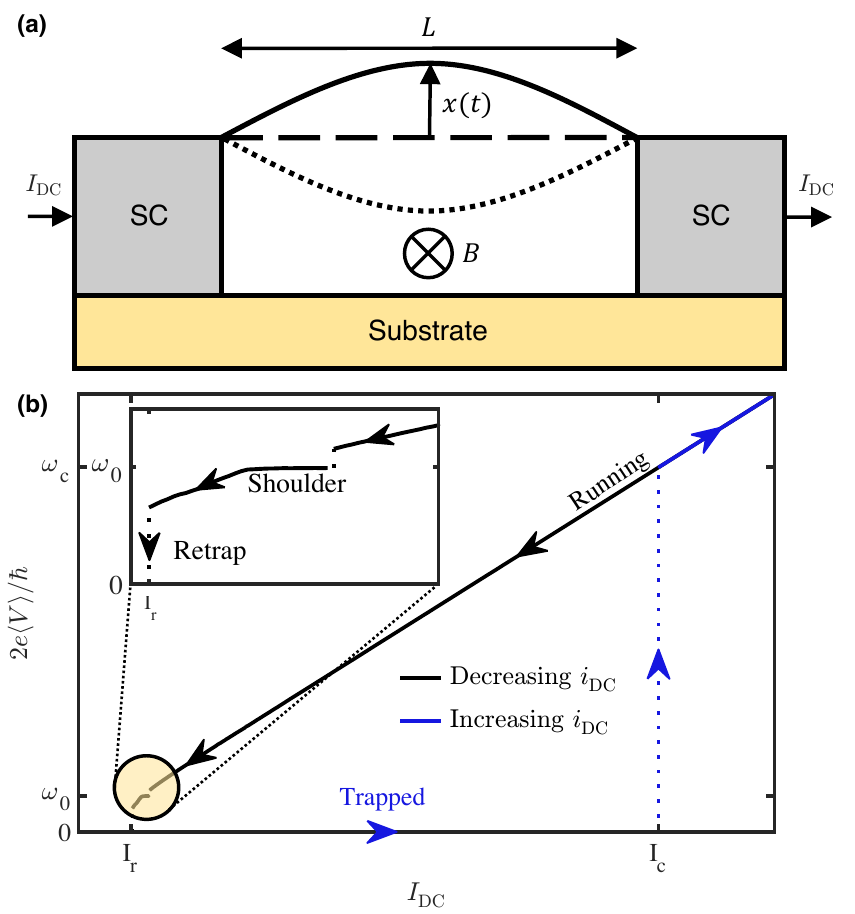}
    \caption{(a) A Josephson junction with a weak-link formed by a mechanical resonator suspended between superconducting contacts, biased with a DC current $I^{}_\text{DC}$. An in-plane magnetic field $B$ mediates the coupling between electronic currents and mechanical oscillations with out-of-plane displacement $x(t)$ and frequency $\omega^{}_0$. 
    (b) DC IVC of an underdamped junction, exhibiting a small mechanically-induced voltage shoulder. This is accessed by first biasing the junction above $I^{}_{\text{c}}$ (blue line), before reducing the bias to achieve resonance at $\expval{V}\approx V^{}_0 = \hbar\omega^{}_0/2e$ (black line) \cite{mcdermott}. Dotted lines correspond to discontinuous transitions between voltage states. For typical setups, on the increasing current path the system jumps to the first finite frequency $\omega^{}_\text{c}\gg\omega^{}_0$. The hysteretic IVC is exploited to achieve resonance in the decreasing current path. 
    Inset: zoom of the IVC around the mechanically-induced shoulder in the decreasing current path.}
    \label{fig:Setup}
\end{centering}
\end{figure}

To address this challenge and unveil electro-mechanical effects in superconducting NEMS under controllable conditions we recently considered a simple setup constituted by a mechanical oscillator suspended between two superconducting contacts, forming a suspended electro-mechanical Josephson junction \cite{sonne,sonne2,mcdermott}, as illustrated in Fig.\,\ref{fig:Setup}(a). 
Josephson predicted that if two superconductors are weakly linked, a supercurrent $I^{}_\text{s} = I^{}_\text{c} \sin\varphi$ will flow between them, where $I^{}_\text{c}$ is the critical current and $\varphi$ is the gauge-invariant superconducting phase difference between the two electrodes \cite{josephson}. %
He further predicted that a voltage V across the junction (either externally imposed by a bias or spontaneously developed by the junction dynamics) would cause the phase to evolve as $d\varphi/dt = 2eV/\hbar$, thus resulting in a high frequency AC supercurrent.
Due to the tiny impedance of the Josephson junction it is very difficult to experimentally maintain a constant voltage bias, and invariably the device will operate in a current-bias mode \cite{likharev}.
With this in mind, we propose a simple setup constituted by a DC current-biased suspended Josephson junction where the coupling between the electronic currents in the weak link and the out-of-plane displacement of the resonator can be controlled by an external in-plane magnetic field via the related Lorentz force. 
In a first analysis of this setup we have shown theoretically \cite{mcdermott} that the excitation of mechanical oscillations of frequency $\omega^{}_0$ in the Josephson weak link results in a Shapiro-like \cite{shapiro} shoulder at a voltage close to $V^{}_0=\hbar\omega^{}_0/2e$ in the DC I-V characteristic (IVC) of a current-biased junction, due to the synchronisation of the oscillating supercurrent and the resonant vibrations (Fig.\,\ref{fig:Setup}(b)).

In the present theoretical paper we unveil for the first time mechanically-induced hysteresis loops in the DC IVC of a current-biased suspended Josephson junction. By traversing the hysteresis loop, a huge voltage plateau appears with a magnetic-field tunable width that can be  larger than half of the critical current. 
This large feature presents the most convincing signature of coupling between superconductors and mechanical resonators to date, providing an experimental advantage for those characterising nano-resonators and searching for signatures of mechanical resonance. 
In contrast to conventional Shapiro steps in the IVC induced by external radiation, here they emerge from spontaneously induced oscillations in the coupled junction-oscillator system.
Energy-sharing between the two coupled systems leads to a number of discrete transitions between the resistive and superconducting states at critical coupling values. We analyse and predict quantitatively these transitions using an analytical framework and energy considerations. The derived analytical expressions allow one to deduce the mechanical quality factor simply by measuring the critical coupling values.

We stress the simplicity of the proposed setup, requiring only DC current bias and voltage measurements to induce and detect mechanical resonances in the suspended Josephson junction. The critical coupling values can be achieved with moderate magnetic fields and temperatures attainable in a commercial cryostat.

\section{Model}
\label{sec:Model}

The set up we consider is depicted in Fig.\,\ref{fig:Setup} (a). A Josephson weak-link with resistance $R$ and capacitance $C$ is biased with a DC current $I^{}_\text{DC}$. 
The weak-link (effectively either one- or two-dimensional in nature) is suspended between two superconducting contacts and acts as a mechanical oscillator with a fundamental flexural mode with displacement amplitude $x$, characterised by an effective mass $M$, length $L$, resonant frequency $\omega^{}_0$ and quality factor $Q$. 
The coupling between the electronic currents flowing through the weak-link and the oscillations is controlled by an in-plane magnetic field $B$.

The experimentally tunable parameters are the bias current and the magnetic field, which we express in dimensionless form as $i^{}_\text{DC} = I^{}_\text{DC}/I^{}_\text{c}$ and $\mu = B/B^{}_0$ respectively. 
Here $I^{}_\text{c}$ is the junction critical current, whilst $B^{}_0 = \sqrt{M/CL^2_{}}$ is the characteristic magnetic field scale of the oscillator.
This set of parameters leads to the length scale $x^{}_0 = B^{}_0 I^{}_\text{c} L / M \omega_0^2$, that corresponds to the displacement at which the mechanical restoring force equals the Lorentz force $B^{}_0 I^{}_\text{c} L$. 
By employing the resistively and capacitively shunted junction (RCSJ) model \cite{stewart} we recently derived the following system of equations \cite{mcdermott} for the gauge-invariant superconducting phase difference $\varphi$ and the dimensionless oscillator displacement $a = x/x^{}_0 - i^{}_\text{DC} \mu$ (measured from the DC offset),
\begin{align}
    & i^{}_{\text{DC}} = \sin \varphi + \beta^{}_1 \dot{\varphi} + \beta^{}_2 \ddot{\varphi} + \mu \ddot{a},
\label{eq:rcsj}\\
    & (1+\mu^2)\ddot{a} + \frac{2}{Q} \dot{a} + a = - \mu \beta^{}_2 \ddot{\varphi}\; .
\label{eq:mech}
\end{align}
Here $\dot{f}$ indicates the derivative of $f$ with respect to the dimensionless time $\tau = \omega^{}_0 t$, and we have defined the dimensionless quantities $\beta^{}_1 = \omega^{}_0/\omega^{}_\text{c}$ (the resonant frequency in units of the junction characteristic frequency $\omega^{}_{\text{c}} = 2eI^{}_{\text{c}}R/\hbar$) and $\beta^{}_2 = \beta_1^2 \beta^{}_\text{c}$ (related to the Stewart-McCumber parameter $\beta^{}_\text{c} = \omega^{}_\text{c} R C$).
In the presence of a finite $\mu$ the mechanical Eq.\,(\ref{eq:mech}) is driven and renormalised by the Lorentz force, whilst the electrical Eq.\,(\ref{eq:rcsj}) is the usual RCSJ model with an additional back-action term due to the motion of the weak-link in a magnetic field. 
\begin{figure}
\begin{centering}
    \includegraphics[width=70mm]{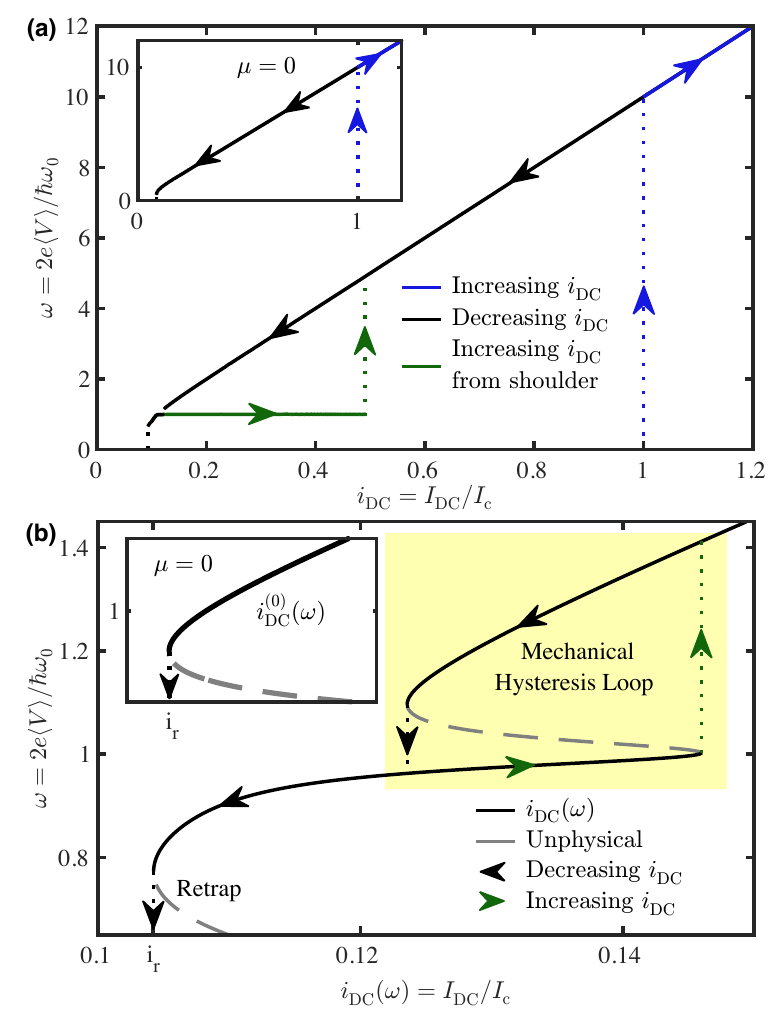}
    \caption{(a) Numerical IVC ($\beta^{}_1 = 0.1$, $\beta^{}_2 = 2$, $\mu = 0.067$, $Q=1000$) produced by increasing $i^{}_{\text{DC}}$ (green line) once locked to the voltage shoulder. For weak coupling a huge voltage plateau of width on the order of $I^{}_\text{c}$ persists until a transition to the Ohmic state, forming a mechanically-induced hysteresis loop. Inset: numerical IVC for $\mu=0$. 
    (b) Analytical $i^{}_{\text{DC}}(\omega)$ plotted with $Q=50$ for illustrative purposes.
    The resonant peak $\delta i^{}_{\text{DC}}(\omega)$ creates a new unphysical region with $\partial i^{}_{\text{DC}}(\omega)/\partial\omega<0$ (grey dashed lines).
    This feature is responsible for the vertical transitions from the Ohmic state to the voltage shoulder (black arrows), and for the larger transition from the plateau maximum to the Ohmic state (green arrows), forming the mechanically-induced hysteresis loop in the DC IVC.
    Inset: The analytical IVC $i_{\text{DC}}^{(0)} (\omega)$, which reproduces the behaviour of an uncoupled junction at $\mu=0$.}
    \label{fig:WeakCoupling}
\end{centering}
\end{figure}

Typical parameters for suspended carbon nanotube systems are \cite{peng,garcia,huttel,laird,moser,jarillo,cleuziou,Mergenthaler} $I^{}_{\text{c}} = 10\si{\nano\ampere}$, $R = 330\si{\ohm}$, $\beta^{}_{\text{c}}=200$, $\omega^{}_0 = 1\si{\giga\hertz}$, $Q = 10^3$, $M = 10^{-20}\si{\kilogram}$ and $L = 1\si{\micro\metre}$, resulting in the derived quantities $\omega^{}_\text{c}=10\si{\giga\hertz},\ V^{}_0 = 0.3\si{\micro\volt},\ B^{}_0 = 10\si{\tesla},\ x^{}_0=10\si{\pico\meter},\ \beta^{}_1 = 0.1$ and $\beta^{}_2 = 2$.
With these parameters the junction operates in the underdamped hysteretic regime ($\beta^{}_\text{c}\gg1$) characterised by an essentially constant voltage $V$ and normal current \cite{likharev}.
When a finite time-average voltage $\expval{V}$ develops across the underdamped junction it drives an AC \emph{electronic supercurrent} of frequency $2e\expval{V}/\hbar$ through the weak link.
This is largely shunted by an AC \emph{displacement current} due to the oscillating electric fields around the junction, while an essentially constant \emph{electronic normal-current} accounts for the remaining DC current bias.
Thus, in the underdamped regime, even with a DC bias current the weak link supports a time-dependent net \emph{electronic (normal and super) current}.
When the AC supercurrent component has a frequency matching the resonance frequency of the normal mode, mechanical oscillations can be activated and amplified controllably via the Lorentz force due to the in-plane magnetic field $B$ \cite{mcdermott}. 

Since typically $\omega^{}_0<\omega^{}_\text{c}$, the mechanical resonance activation can only occur by exploiting the hysteresis loop indicated in Fig.\,\ref{fig:Setup}(b).
Firstly the running (Ohmic) state is entered by driving $i^{}_\text{DC}>1$, yielding finite voltages corresponding to frequencies greater than $\omega^{}_\text{c}$ (blue line).
Then, lower frequencies can be reached by progressively reducing $i^{}_\text{DC}$ (black line) until mechanical activation occurs at the dimensionless frequency $\omega=\expval{\dot{\varphi}}=\expval{V}/V^{}_0\approx 1$ (note that the junction must be suitably underdamped such that retrapping does not occur before resonance is achieved).
This mechanical resonance manifests itself as a voltage shoulder in the DC IVC, as discussed in \cite{mcdermott}.
Once the shoulder is entered, reducing $i^{}_\text{DC}$ further leads to retrapping at $i^{}_\text{r}=I^{}_\text{r}/I^{}_\text{c}\approx 4\beta^{}_1/\pi\sqrt{\beta^{}_2}$, as shown in the inset in Fig.\,\ref{fig:Setup}(b).


\section{Weak-Coupling}
\label{sec:WeakCoupling}



A set of natural questions arises: can one extend the size of the shoulder to enter a full mechanically-induced plateau by \emph{increasing} $i^{}_{\text{DC}}$ once the mechanical resonance is reached? What is the fate of this plateau for larger coupling?  
To address these questions we first performed a numerical analysis of Eqs.\,(\ref{eq:rcsj}) and (\ref{eq:mech}) with finite $\mu$.
The result is staggering: after entering the shoulder in the DC IVC, by increasing $i_\text{DC}^{}$ one observes a massive mechanically-induced voltage plateau at $\omega\approx 1$, with width on the order of half of the junction critical current (continuous green line in Fig.\,\ref{fig:WeakCoupling}(a)). This plateau is reminiscent of the  Shapiro steps induced by external radiation, while here the AC signal is provided by the spontaneously induced mechanical vibrations. For small values of the coupling $\mu$ (to be quantified in Sec.\,\ref{sec:StrongCoupling}), the plateau extends until a sudden transition occurs to the Ohmic state (dashed green line in Fig.\,\ref{fig:WeakCoupling}(a)), forming a massive mechanically-induced hysteresis loop in the IVC. This feature is to be contrasted with the IVC of an underdamped junction in the absence of mechanical oscillations ($\mu=0$), as shown in the inset in Fig.\,\ref{fig:WeakCoupling}(a).
 
To understand this behaviour and assess the size of the loop as a function of the coupling strength, we study analytically the relation between $i_\text{DC}$ and $\omega$. In order to derive $i_\text{DC}(\omega)$ we employ the following ansatzes for the gauge-invariant phase difference $\varphi$ and the oscillator displacement $a$ in the underdamped regime \cite{mcdermott}
\begin{align}
    & \varphi = \varphi^{}_0+\omega\tau -\frac{g}{\omega}\cos(\omega\tau),
\label{eq:phase_ansatz}\\
    & a = A\sin(\omega\tau+\theta)\; .
\label{eq:mech_ansatz}
\end{align}
Here $g\ll\omega$ parametrises a small mechanically-induced voltage fluctuation, $A$ is the oscillator amplitude and $\theta$ is a phase delay. Inserting these ansatzes into the equations of motion (\ref{eq:rcsj}) and (\ref{eq:mech}) we deduce the current function $i^{}_\text{DC}(\omega,\mu) = i_\text{DC}^{(0)}(\omega) + \delta i^{}_\text{DC} (\omega,\mu)$ that is plotted in Fig.\,\ref{fig:WeakCoupling}(b). This is conveniently decomposed as the sum of a $\mu$-independent smooth profile $i_\text{DC}^{(0)}(\omega)$ and a $\mu$-dependent peak $\delta i^{}_\text{DC} (\omega,\mu)$ centred around $\omega \approx 1$. The uncoupled IV characteristic ($\mu=0$) is given by
\begin{equation}
	i_\text{DC}^{(0)}(\omega) = \beta_1 \left(\omega + \frac{1}{2\omega(\beta_1^2+\beta_2^2 \omega^2)}\right),
    \label{idc0eq}
\end{equation}
which is plotted in the inset in Fig.\,\ref{fig:WeakCoupling}(b). 
Regions with $\partial i^{}_\text{DC}/\partial \omega < 0$ (indicated as grey dashed lines) are regarded as unphysical, so that the physical part of this curve ranges from a minimum of $\omega \approx (3/2)^{1/4}_{}/\sqrt{\beta^{}_2}$ corresponding to the uncoupled retrapping frequency (\textit{cf}. the common result $4/\pi\sqrt{\beta^{}_2}$), to the ohmic behaviour $i^{(0)}_\text{DC}(\omega) \to \beta^{}_1 \omega$ for $\omega\gg\beta^{-1/2}_{2}$. 

At finite $\mu$ the mechanical peak $\delta i^{}_\text{DC}(\omega,\mu)$ is manifested as a large resonance centred at $\omega\approx 1$ in the DC IVC (Fig. \ref{fig:WeakCoupling}(b)). 
The physical portion of this peak ($\partial i^{}_\text{DC}/\partial \omega > 0$) is responsible for the large voltage plateau in Fig.\,\ref{fig:WeakCoupling}(a) that is explored by increasing the current bias (green arrows in Fig. \ref{fig:WeakCoupling}(b)).
%
%
If one increases the current bias past the peak in $\delta i^{}_\text{DC}(\omega,\mu)$, to remain in a physical state the system makes a vertical voltage transition to the Ohmic branch (green dotted line), forming the mechanical hysteresis loop highlighted by the yellow box in Fig. \ref{fig:WeakCoupling}(b).
The shape and width of the mechanically-induced plateau is entirely determined by the function $\delta i^{}_\text{DC}(\omega,\mu)$ (see full expression in the Supplementary Material). The essential features of the plateau are captured by expanding $\delta i^{}_\text{DC}(\omega,\mu)$ about $\omega\approx1$. Here the function takes the form of a Fano-resonance  
\begin{equation}
    \delta i^{}_\text{DC}(\omega,\mu) \approx \frac{\mu^2 q_{}^3}{(1+q_{}^2)^2\beta_{2}^{}}\left[\frac{(\omega^2-\omega_\text{Fano}^2)+\frac{q^2-1}{q\Gamma}\left(\frac{\Gamma}{2}\right)^2}{(\omega^2-\omega_\text{Fano}^2)^2+\left(\frac{\Gamma}{2}\right)^2}\right],
    \label{eq:fano}
\end{equation}
where the Fano-frequency is $\omega_\text{Fano}^2 = 1-\frac{\mu^2 }{1+q_{}^2}$, the
linewidth $\Gamma = \frac{2\mu^2 q}{1+q_{}^2}+\frac{4}{Q}$ and  $q=\beta^{}_2/\beta_1^{}$.
In the limit $q \to \infty$ the Fano formula reduces to the usual Breit-Wigner (Lorentzian) expression. 
The peak frequency, corresponding to the voltage plateau, is equal to the value $\omega^{}_\text{max} = \sqrt{\omega_\text{Fano}^2+\Gamma/2q} = \sqrt{1+2/q Q}$ which is independent of $\mu$ and extremely close to the bare resonance frequency
$\omega = 1$.
%
%
The independence of the plateau frequency $\omega^{}_\text{max}$ on $\mu$ is important for metrology applications.
The width of the voltage plateau in the DC IVC is determined by the height of the peak $\Delta i^\text{max}_\text{DC}(\mu)=\delta i^{}_\text{DC}(\omega^{}_\text{max},\mu)$
\begin{equation}\label{eq:WeakCouplingPlateauWidth}
    \Delta i_{\text{DC}}^{\text{max}}(\mu)= \frac{\mu^2 q^4}{(1+q_{}^2)^2\Gamma\beta_{2}^{}}.
\end{equation}
The plateau width thus increases quadratically with $\mu$ until a strong coupling regime is entered, where this analysis breaks down, as detailed below.
Analytical insight in the mechanical oscillation amplitude $A$ is provided in the Supplementary Material, while here we concentrate on the fate of the IVC in the strong coupling regime.


\section{Strong-Coupling}\label{sec:StrongCoupling}

\begin{figure*}[htb!]
\begin{centering}
    \includegraphics[width=170mm]{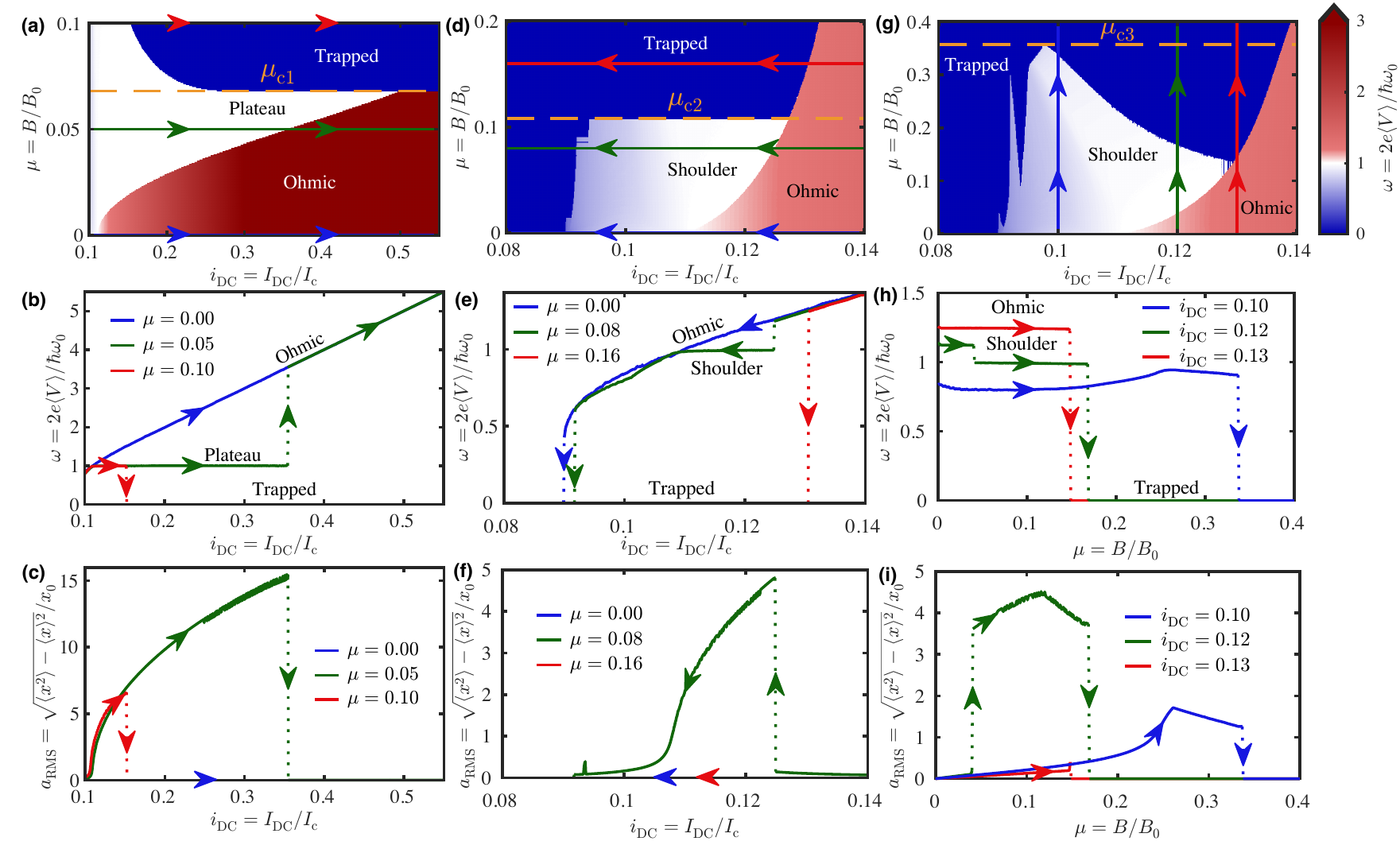}
    \caption{(a), (d), (g) Colour plots of the DC dimensionless voltage $\omega$ as a function of $i^{}_{\text{DC}}$ and $\mu$, obtained numerically for the parameters $\beta^{}_1 = 0.1$, $\beta^{}_2 = 2$ and $Q =10^3$ via the hysteretic procedures outlined below.
    (a) Increasing $i^{}_\text{DC}$ from the voltage shoulder at fixed $\mu$. For weak coupling the width of the voltage plateau (white region) scales as $\mu^2_{}$, and the system returns to the Ohmic state (red region) at the plateau edge. For strong coupling $\mu>\mu^{}_\text{c1}$ (gold dashed line) the system undergoes mechanically-induced retrapping to the superconducting state (blue region), quenching the plateau width.
    (b) Cuts at fixed $\mu$ corresponding to the arrow lines in panel (a). For $\mu=0.1>\mu^{}_\text{c1}$ (red line) the plateau width is reduced with respect to $\mu=0.05<\mu^{}_\text{c1}$ (green line), and the mechanically-induced hysteresis loop is broken.
    (c) Mechanical amplitude $a^{}_\text{RMS}$ corresponding to the cuts in panel (b). The RMS mechanical amplitude increases monotonically with $i^{}_\text{DC}$ along the plateau, and is curtailed when transitions to the Ohmic or trapped states occur.
    (d) A similar colour plot to (a), produced by reducing the current bias on the Ohmic branch until the voltage shoulder is entered. For $\mu<\mu^{}_\text{c2}$ (gold dashed line) the system enters the shoulder state (green line), whilst for $\mu>\mu^{}_\text{c2}$ the voltage shoulder becomes inaccessible and premature-retrapping occurs (red line). This is highlighted in the cuts at fixed $\mu$ in panel (e).
    (f) Mechanical amplitude $a^{}_\text{RMS}$ corresponding to the cuts in panel (e). For $\mu>\mu^{}_\text{c2}$ (red line) no persistent mechanical oscillations are excited.
    (g) Here the system is prepared in the Ohmic state (for $i^{}_\text{DC}>i^{}_\text{r}$) at $\mu=0$, before increasing $\mu$ at fixed $i^{}_\text{DC}$. The voltage shoulder can be entered via this procedure and persists to large values of $\mu$ before retrapping occurs. The maximum value of $\mu$ at which shoulder states exist is $\mu^{}_\text{c3}$ (gold dashed line).
    (h) Cuts of (g) at fixed $i^{}_\text{DC}$ corresponding to the arrow lines in panel (g). The cut at $i^{}_\text{DC}=0.12$ (green line) clearly demonstrates voltage shoulder entry via this procedure.
    (i) Mechanical amplitude $a^{}_\text{RMS}$ corresponding to the cuts in panel (h), showing the excitation of mechanical oscillations on the voltage shoulder.}
    \label{fig:StrongCouplingNumerics}
\end{centering}
\end{figure*}

The voltage plateau in our setup emerges from spontaneously induced oscillations in the coupled junction-oscillator system. As such, the electronic and mechanical systems are subject to energy exchange, which can cause \textit{mechanically-induced retrapping} events in the IVC at large coupling, as illustrated in Fig.\,\ref{fig:StrongCouplingNumerics}. 

Fig.\,\ref{fig:StrongCouplingNumerics}(a) shows the dimensionless DC voltage produced by increasing $i^{}_\text{DC}$ from the shoulder (keeping $\mu$ fixed) for a range of $\mu$.
For $\mu$ smaller than the first critical coupling $\mu^{}_\text{c1}$ ($\mu^{}_\text{c1}\approx0.07$ for the parameters used in the figure) the width of the plateau increases quadratically with $\mu$, as predicted by Eq.\,\ref{eq:WeakCouplingPlateauWidth}.
Increasing $i^{}_\text{DC}$ beyond the plateau edge induces a transition to the Ohmic state, forming the hysteresis loop.

Our analysis shows that the picture above breaks down for $\mu>\mu^{}_\text{c1}$. 
In stark contrast to the transition to the Ohmic state, here the system undergoes a sudden retrapping to the superconducting state from the voltage plateau due to large energy transfer from the electronic to the mechanical system.
This has the effect of heavily reducing the plateau width and breaking the mechanical hysteresis loop (as shown in the cuts at fixed $\mu$ in Fig.\,\ref{fig:StrongCouplingNumerics}(b)).
Fig.\,\ref{fig:StrongCouplingNumerics}(c) shows the amplitude of the mechanical oscillations corresponding to the cuts at constant $\mu$ in Fig.\,\ref{fig:StrongCouplingNumerics}(b)). Oscillations are amplified while increasing $i^{}_\text{DC}$ on the voltage plateau until they are suddenly quenched at the plateau edge where retrapping occurs. 

This is not the only example of a mechanically-induced retrapping in the DC IVC. 
Indeed, a similar scenario occurs even when reducing $i^{}_\text{DC}$
from the Ohmic state in an attempt to access the mechanical voltage shoulder, as shown in Fig.\,\ref{fig:StrongCouplingNumerics}(d).
For $\mu$ less than the second critical coupling $\mu^{}_\text{c2}$ ($\mu^{}_\text{c2}\approx 0.11$ for the chosen parameters) the voltage shoulder is accessible and mechanical oscillations are excited, whilst for $\mu>\mu^{}_\text{c2}$ premature retrapping occurs to the superconducting state \cite{mcdermott}. 
The cuts at fixed $\mu$ of the dimensionless DC voltage in Fig.\,\ref{fig:StrongCouplingNumerics}(e) and the corresponding mechanical amplitude in Fig.\,\ref{fig:StrongCouplingNumerics}(f) show that the premature retrapping prevents the system from entering the shoulder state, resulting in the suppression of any mechanical oscillations and in the increase of the retrapping current.

It is important to point out that the IVC of the system changes if one instead performed the measurement by varying $\mu$ at a fixed current bias, as shown in Fig.\,\ref{fig:StrongCouplingNumerics}(g).
If $i^{}_\text{DC}<0.09$ (for the parameters in the figure) the system is in the superconducting state irrespective of the value of $\mu$. 
If instead the system is prepared in the Ohmic state for $0.09 \lesssim i^{}_\text{DC}\lesssim 0.13$ and $\mu$ is increased from zero, a small voltage shoulder is accessible as shown in the cuts in Fig.\,\ref{fig:StrongCouplingNumerics}(h).
The resulting mechanical amplitude in Fig.\,\ref{fig:StrongCouplingNumerics}(i) confirms the excitation of mechanical oscillations.
The voltage shoulder generated by this procedure persists to values of $\mu$ larger than $\mu^{}_\text{c1}$ and $\mu^{}_\text{c2}$.
The maximum coupling at which shoulder states exist is labelled the third critical coupling $\mu^{}_\text{c3}\approx 0.36$.
We point out that our numerical solution exhibits self-similar structures in the $(i^{}_\text{DC},\mu)$ phase space, manifesting themselves as sharp tongues at $i^{}_\text{DC}\approx 0.091$ and $0.093$ in Fig.\,\ref{fig:StrongCouplingNumerics}(g).
These correspond to micro-shoulders in the IVC at the fractional frequencies $\omega\approx 1/2$ and $2/3$ respectively, due to the nonlinearity of the Josephson element.
Accessing these micro-shoulders while decreasing $i^{}_\text{DC}$ (\textit{e.g.} the green line in Fig.\,\ref{fig:StrongCouplingNumerics}(d)) corresponds to the spike of mechanical amplitude observed at $i^{}_\text{DC}\approx0.093$ in Fig.\,\ref{fig:StrongCouplingNumerics}(f).

To understand these retrappings analytically beyond the numerical approach above, we examine in detail the energy of the system.
The dimensionless energy  stored in the electronic system  is equal to $E^{}_{\varphi} = \frac{1}{2}\beta^{}_2 \dot{\varphi}^2 - \cos \varphi$ (expressed in units of the Josephson energy $E^{}_\text{J}=\hbar I^{}_\text{c}/2e$, see \cite{mcdermott}). 
While $E^{}_\varphi$ is oscillating in time, we find that a very effective condition for retrapping is obtained when its time average falls below the maximum of the Josephson potential $\cos \varphi$ \textit{i.e.} $\expval{E^{}_\varphi} < 1$. 
%
%
As the total energy is shared between the electronic and mechanical systems, the excitation of mechanical oscillations increases the mechanical energy at the expense of the electronic energy.
Above a threshold coupling we may reach a point where the electronic system has too little energy to escape from a local minimum in the Josephson potential, causing a mechanically-induced retrapping to the superconducting state $\omega=\expval{\dot{\varphi}}=0$.
As such, the activation of mechanical oscillations can be viewed as an additional damping mechanism for the electronic degrees of freedom.

\begin{figure*}[htb!]
\begin{centering}
    \includegraphics[width=180mm]{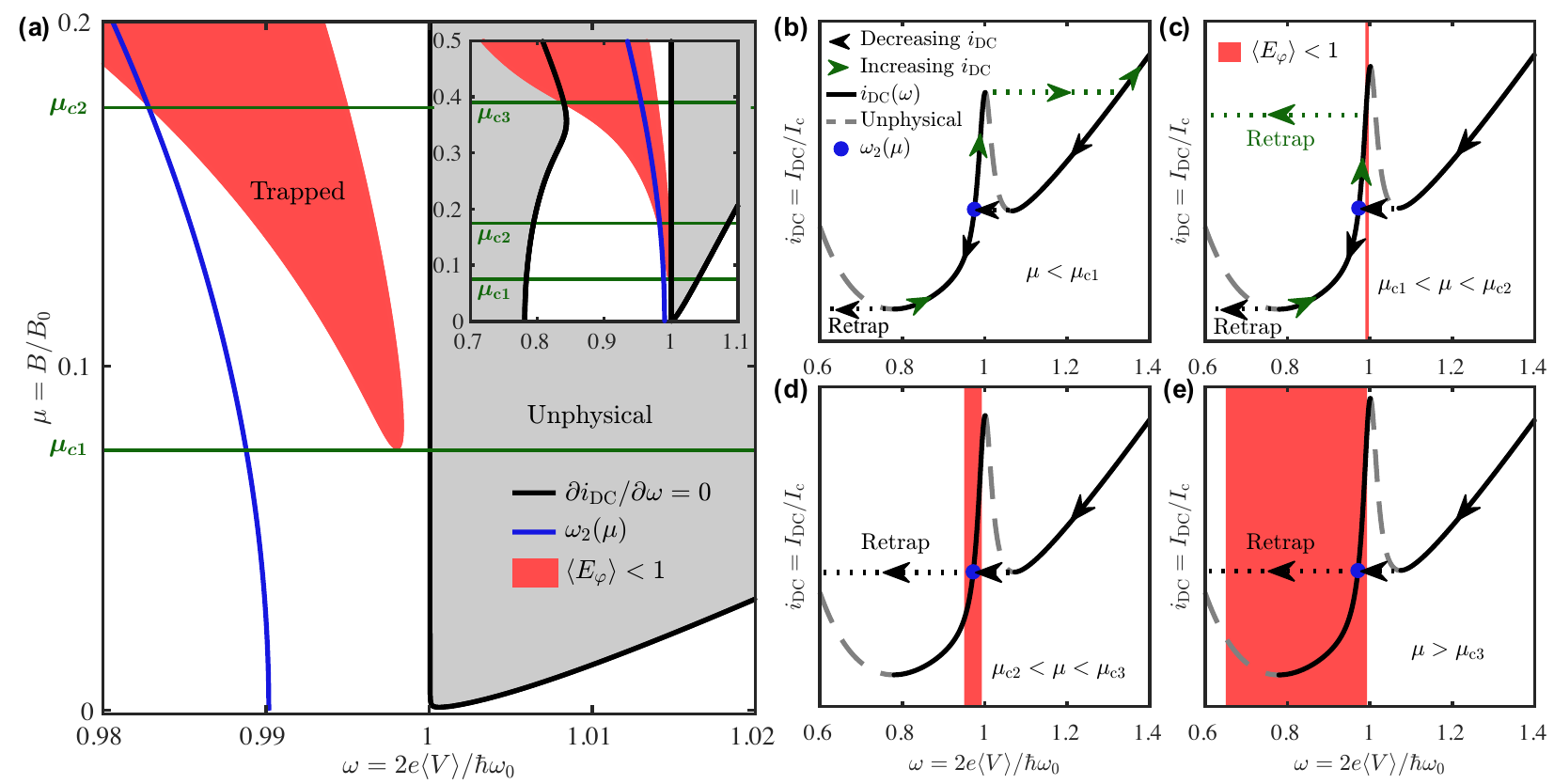}
    \caption{(a) Contour plot illustrating the phase space $(\omega,\mu)$ for parameters $\beta^{}_1=0.1$, $\beta^{}_2=2$ and $Q=10^3$. Black lines indicate $\partial i^{}_{\text{DC}}/\partial\omega=0$, separating physical/unphysical solutions (white/grey shading, respectively). 
    The blue line indicates the frequency $\omega^{}_2(\mu)$ at which the system attempts to enter the voltage shoulder.
    The first critical coupling $\mu^{}_{\text{c1}}$ is defined as the smallest value of $\mu$ where the retrapping region $\expval{E^{}_{\varphi}}<1$ (red shading) exists.
    The second critical coupling $\mu^{}_\text{c2}$ is defined as the value of $\mu$ at which  $\omega^{}_2(\mu)$ enters the trapped region, preventing the system from entering the shoulder state by reducing $i^{}_\text{DC}$.
    Inset: a plot of the phase space for a larger span of $\omega$ and $\mu$. The third critical coupling $\mu^{}_\text{c3}$ is defined as the value of $\mu$ where the entire voltage shoulder is engulfed by the trapped region. 
    (b) For $\mu<\mu^{}_\text{c1}$ all states in the voltage plateau may be accessed via the decreasing (black arrows) and increasing (green arrows) current paths, giving rise to the mechanically-induced hysteresis loop (see main text for further details).
    (c) For $\mu^{}_{\text{c1}}<\mu<\mu^{}_{\text{c2}}$ a portion of the mechanically-induced plateau becomes trapped (red stripe), causing retrapping in the increasing current path (green arrows).
    (d) For $\mu^{}_{\text{c2}}<\mu<\mu^{}_{\text{c3}}$ the frequency $\omega^{}_2(\mu)$ becomes trapped, and the system directly retraps whilst attempting to enter the voltage shoulder from the Ohmic branch on the decreasing current path.
    (e) For $\mu>\mu^{}_{\text{c3}}$ all states on the voltage shoulder become trapped.
    }
    \label{fig:StrongCouplingAnalytics}
\end{centering}
\end{figure*}

In order to predict analytically when the retrapping occurs, the ansatzes (Eqs.\,\ref{eq:phase_ansatz} and \ref{eq:mech_ansatz}) are substituted into the electronic energy to obtain an expression for $\expval{E^{}_{\varphi}(\omega,\mu)}$ (in-depth derivations of these results are available in the Supplementary Material).
The retrapping region $\expval{E^{}_\varphi(\omega,\mu)}<1$ is plotted in red in Fig.\,\ref{fig:StrongCouplingAnalytics}.

In Fig.\,\ref{fig:StrongCouplingAnalytics}(a) unphysical states ($\partial i^{}_\text{DC}/\partial\omega<0$) are indicated by grey shading in the phase space ($\omega,\mu$), while the white regions indicate viable states. At this point, we just use the form of the current function $i^{}_\text{DC}(\omega,\mu)$ to determine the frequency $\omega$ corresponding to a given DC bias $i^{}_\text{DC}$ and coupling $\mu$. If this $\omega$ falls in the region $\expval{E^{}_\varphi(\omega,\mu)}<1$, retrapping occurs, otherwise the system stays in the plateau or transitions back to the ohmic curve.

For $\mu<\mu^{}_\text{c1}$ no trapped states exist, and the IVC is determined by the weak coupling dynamics, summarised by the analytical IVC in Fig.\,\ref{fig:WeakCoupling}(b), and corresponding to the qualitative sketch in Fig.\,\ref{fig:StrongCouplingAnalytics}(b).
The shoulder is accessed by decreasing $i_{\text{DC}}$ from the Ohmic branch until one reaches the local minimum with $\partial i^{}_{\text{DC}}/\partial\omega=0$ at $\omega\approx 1.05$ in Fig.\,\ref{fig:StrongCouplingAnalytics}(b).
By further reducing $i^{}_\text{DC}$ a sudden transition to the shoulder occurs, to the frequency $\omega^{}_2$ (blue line in Fig.\,\ref{fig:StrongCouplingAnalytics}(a) and blue dot in Fig.\,\ref{fig:StrongCouplingAnalytics}(b)).
Decreasing $i^{}_{\text{DC}}$ from here explores the shoulder of the voltage resonance, before the system retraps at the global minimum of the DC current ($\partial i^{}_{\text{DC}}/\partial\omega=0$) at $\omega\approx 0.8$.
If instead $i^{}_{\text{DC}}$ is increased from the frequency $\omega^{}_2$, one explores the whole voltage plateau until the peak of the voltage resonance at $\omega^{}_\text{max}$, at which point the system transitions back to the Ohmic state (green dotted lines in Fig.\,\ref{fig:StrongCouplingNumerics}(b) and Fig.\,\ref{fig:StrongCouplingAnalytics}(b)), forming the hysteresis loop.

For $\mu>\mu^{}_\text{c1}$ a trapped region appears in the phase space $(\omega,\mu)$, for a window of frequencies less than $\omega^{}_\text{max}$.
If the frequency of the system in the IVC enters this trapped region, the system will mechanically-retrap to the superconducting state.
This is demonstrated in Fig.\,\ref{fig:StrongCouplingAnalytics}(c), where the system undergoes sudden mechanical retrapping whilst increasing $i^{}_\text{DC}$ along the voltage plateau.
This mechanical retrapping (green dotted line in Fig.\,\ref{fig:StrongCouplingAnalytics}(c)) causes a truncation of the voltage plateau and the breaking of the hysteresis loop (corresponding to the red line in Fig.\,\ref{fig:StrongCouplingNumerics}(b)).
%
%
An analytical expression for $\mu^{}_\text{c1}$ can be derived by performing an expansion of $\expval{E^{}_\varphi}$ in the small parameters $\delta\omega^2_{}=1-\omega^2_{},\ 1/Q$ and $\mu^2_{}$, yielding a minimum of the trapped contour $\expval{E^{}_\varphi}=1$ at 
\begin{equation}
    \mu^{}_{\text{c}1} \approx \frac{1}{\sqrt{Q}}
    \sqrt{\frac{4(\beta^{}_2-2)\beta^{}_2+6}{\sqrt{1-2\beta^{}_2(\beta^{}_2-2)}}}\; .
    \label{eq:muc1}
\end{equation}
Combining this critical value with the expression for $\Delta i^\text{max}_\text{DC}(\mu)$ in Eq.\,\ref{eq:WeakCouplingPlateauWidth} we obtain the maximum size of the voltage plateau
\begin{equation}
    \Delta i^\text{max}_\text{DC}(\mu^{}_{\text{c}1}) \approx \frac{2\beta_2(\beta_2-2)+3}{2\beta_2\sqrt{1-2\beta_2(\beta_2-2)}}
    \label{eq:shapiromax}
\end{equation}
that is independent of $Q$. For the 
parameters chosen here, this corresponds to a plateau width $\Delta i^\text{max}_\text{DC}=3/4$. 

By progressively increasing $\mu$ above $\mu^{}_\text{c1}$, one reaches a critical value of the coupling $\mu^{}_\text{c2}$ such that even the frequency $\omega^{}_2$ is engulfed by the trapped region, as shown in Fig.\,\ref{fig:StrongCouplingAnalytics}(a).
For $\mu>\mu^{}_\text{c2}$ the transition from the Ohmic branch to the voltage shoulder at $\omega^{}_2$ is thus forbidden, and the system directly retraps to the superconducting state (as indicated in Fig.\,\ref{fig:StrongCouplingAnalytics}(d), corresponding to the red line in the numerical IVC in Fig.\,\ref{fig:StrongCouplingNumerics}(e)).
In order to calculate $\mu^{}_\text{c2}$, we use an approximation of $\omega^{}_2(\mu)$ (see the Supplementary Material) and calculate its intersection with the contour $\expval{E^{}_\varphi}=1$, leading to
\begin{equation}
    \mu^{}_{\text{c}2} \approx \frac{1}{\sqrt{Q}} \sqrt{\frac{2\beta_2}{\beta_1}\left(\frac{2}{\sqrt{1-2\beta_2(\beta_2-2)}}-1\right)}\; .
    \label{eq:muc2}
\end{equation}
According to Eqs.\,(\ref{eq:muc1}) and (\ref{eq:muc2}) both $\mu^{}_{\text{c}1}$ and $\mu^{}_{\text{c}2}$ scale as $1/\sqrt{Q}$, and they both diverge as $2\beta_2 (\beta_2-2) \to 1$, yielding a critical value $\beta^{}_{2,\text{crit}} \approx 1+\sqrt{3/2} \approx 2.225$ above which the system can no longer retrap.
Numerical investigation shows that such a critical value of $\beta^{}_2$ does indeed exist, but is larger than predicted analytically, with a value $\beta^{\mathrm {num}}_{2,\text{crit}}\sim 5$. 
This quantitative disagreement arises due to the assumptions of $\mu^2 \ll 1$ and small $\varphi$ oscillations being violated in this regime.

Even for $\mu>\mu^{}_\text{c2}$, physically allowed states may exist between the global minimum of the curve $i^{}_\text{DC}(\omega)$  (located at $\omega\approx0.8$ in Fig.\,\ref{fig:StrongCouplingAnalytics}(d)) and the retrapping region, but as $\omega^{}_2$ is unstable these states cannot be accessed simply by reducing $i^{}_\text{DC}$ from the Ohmic branch.
To access these states in the IVC, one may prepare the system at a frequency in this window at $\mu=0$ before increasing the coupling, corresponding to the numerical procedure in Fig.\,\ref{fig:StrongCouplingNumerics}(g-i).
As long as these states do not overlap with the red retrapping region, the system may access them even for $\mu>\mu^{}_\text{c2}$.
Indeed, the upper cutoff coupling $\mu^{}_\text{c3}$ is defined as the point where all physical states on the mechanical branch are engulfed by the retrapping region, as indicated in Fig.\,\ref{fig:StrongCouplingAnalytics}(e).
In the zoomed out inset in Fig.\,\ref{fig:StrongCouplingAnalytics}(a), this corresponds to the intersection of the retrapping contour with the global minimum of the $i^{}_\text{DC}(\omega,\mu)$ curve with $\partial i^{}_\text{DC}/\partial\omega=0$.
Since this minimum has a weak dependence on $\mu$, we may approximate it by the uncoupled value $\partial i_\text{DC}^{(0)}/ \partial \omega = 0$ \textit{i.e.} $\omega \approx (3/2)^{1/4}/\sqrt{\beta^{}_2}$. 
Substituting this into the retrapping contour $\expval{ E^{}_{\varphi}}=1$ and solving for $\mu$ yields
\begin{equation}
    \mu_{\text{c}3} \approx \frac{1}{2}\sqrt{\beta_2-\sqrt{3/2}}\; ,
    \label{eq:muc3}
\end{equation}
which predicts that $\mu^{}_{\text{c}3}$ is independent of $Q$.
This is in excellent agreement with numerical simulations (see Supplementary Material).
%



\section{Discussion and Conclusions}


In this paper we have shown how spontaneously-induced mechanical oscillations manifest themselves in the DC IVC of a suspended Josephson junction, under purely DC current bias.
The coupling ($\mu$) between electronic and mechanical degrees of freedom can be controlled by means of an external in-plane magnetic field.
We show that a mechanically induced hysteresis loop arises in the DC IVC, providing clear evidence of the activation and transduction of mechanical oscillations, even in the notoriously difficult GHz regime. 
We predict the appearance of a huge mechanically-induced voltage plateau in the IVC, of a width comparable to the junction critical current.
The direct measurement of the plateau voltage $V^{}_\text{plat}$ directly relates to the mechanical resonance frequency $\omega^{}_0$ 
\begin{equation}\label{eq:Vplat_expression}
    V^{}_\text{plat} =\frac{\hbar\omega^{}_0}{2e} \sqrt{1+\frac{2\beta^{}_1}{\beta_2^{}Q}},
\end{equation}
which is only marginally renormalised from the bare resonance frequency for typical suspended devices of large quality factor $Q$. 
The latter may be characterised by determining the critical coupling $\mu^{}_\text{c1}$ at which the large plateau is quenched.
This results in a value 
\begin{equation}\label{eq:Q_expression}
    Q \approx \frac{1}{\mu^{2}_{\text{c}1}}
    \frac{4(\beta^{}_2-2)\beta^{}_2+6}{\sqrt{1-2\beta^{}_2(\beta^{}_2-2)}}\; .
\end{equation}
In the Supplementary Material we provide full details of the procedure to experimentally deduce all the parameters in our model.
The large size of the voltage plateau combined with the independence of $V^{}_\text{plat}$ on the value of the electromechanical coupling $\mu$ may be exploited in metrology applications such as force and mass sensing.
Despite the invariance of the DC voltage across the plateau, we show that the mechanical amplitude can be amplified continuously by increasing the DC bias current.

Our analytical investigation of the energy sharing between the electronic and the mechanical systems allows us to explain all the retrappings occurring in our system, and provides reliable estimates for the mechanical quality factor. 
Discrepancies between simulations and analytics can arise due to the breakdown of some of our approximations. 
In particular, the ratio $g/\omega$ becomes large at high current values on the plateau and near the upper critical coupling $\mu^{}_{\text{c}3}$, and the single-frequency ansatz fails to capture the self-similar features in Figs.\,\ref{fig:StrongCouplingNumerics}(d) and (g). 
These features occupy only a small region in the phase space, and we expect them to be challenging to detect experimentally.

One of the most important aspects of this work is the relative simplicity of the proposed experimental setup, that can be realised by both one and two-dimensional suspended weak links such as carbon nanotubes and graphene.
%
%
The typical magnetic fields involved in the proposed experiments are of order $B_{\text{c}1}\sim700\si{\milli\tesla}$ (corresponding to $\mu_{\text{c}1}$) for a very moderate $Q=10^3_{}$.
This is easily attainable for nano-scale devices, and can be further reduced using state of the art resonators with $Q>10^6$ \cite{moser}.

While the results discussed in the text are at zero temperature, the effect of finite temperature $T$ was studied via the addition of a Johnson-Nyquist noise current to Eq.\,\ref{eq:rcsj} \cite{Lee}.
Performing stochastic simulations at finite $T$, we find that the thermally increased retrapping current \cite{Kautz} will cause the underdamped condition $I^{}_\text{r}(T)<\beta^{}_1 I{}_\text{c}$ to break down at a critical value of $T$, above which accessing the mechanical shoulder and plateau is not feasible. 
For example, a typical carbon nanotube resonator with $\omega^{}_0=2\si{\giga\hertz}$ permits plateau entry up to $T\approx 20 \si{\milli\kelvin}$, a temperature attainable in a commercial dilution refrigerator.
This may be further improved by using higher frequency mechanical resonators with resonant frequency of several hundred $\si{\giga\hertz}$  \cite{island}.
Once the underdamped condition is fulfilled, finite $T$ causes premature switching from the mechanically induced voltage plateau to the Ohmic state, resulting in a reduced plateau and hysteresis loop; this may be countered via the use of high $Q$ resonators \cite{moser}.

\section*{ACKNOWLEDGMENTS}
\noindent This work was supported by the Engineering and Physical Sciences Research Council of the United Kingdom through the EPSRC Centre for Doctoral Training in Metamaterials (grant number EP/L015331/1), the EU H2020-MSCA-RISE project DiSeTCom (Project No. 823728) as well as by the Royal Society (Grant No. IEC/R2/192166). 




\end{document}


\section{Analytics}

\subsection{Ansatz}

In this section we discuss the analytical approach used to deduce current-voltage relation $i^{}_\text{DC}(\omega ,\mu)$, finally leading to the DC IVC of the suspended Josephson junction.  
The dimensionless equations of motion 
\begin{equation}\label{eq:supp:rcsj}
    i^{}_{\text{DC}} = \sin \varphi + \beta^{}_1 \dot{\varphi} + \beta^{}_2 \ddot{\varphi} + \mu \ddot{a},
\end{equation}
\begin{equation}\label{eq:supp:mech}
    (1+\mu_{}^2)\ddot{a} + \frac{2}{Q} \dot{a} + a = - \mu \beta^{}_2 \ddot{\varphi}
\end{equation}
have been derived in our previous work\footnote[1]{T. McDermott, H.-Y. Deng, A. Isaccson, E. Mariani, \href{https://doi.org/10.1103/PhysRevB.97.014526}{Phys. Rev. B \textbf{97}, 014526 (2018)}.}.
%
Expressions for the mechanical amplitude and IVC were derived by substitution of the ansatzes
\begin{equation}\label{eq:supp:phaseansatz}
    \varphi = \varphi^{}_0+\omega\tau -\frac{g}{\omega}\cos(\omega\tau),
\end{equation}
\begin{equation}\label{eq:supp:dissplacementansatz}
    a = A\sin(\omega\tau+\theta)
\end{equation}
into Eqs.\,\ref{eq:supp:rcsj} and \ref{eq:supp:mech} in the limit of small phase fluctuations $g/\omega\ll 1$ usually associated with underdamped Josephson junctions. The validity of this assumption will be discussed below.
This resulted in the expressions
\begin{equation}\label{eq:supp:IVC}
    i^{}_\text{DC} = \beta^{}_1\omega+\frac{g^2}{2\omega} \left(\beta^{}_1+\mu^2\beta_2\omega^3\frac{2\omega}{Q\kappa}\right),
\end{equation}
\begin{equation}\label{eq:supp:A}
    A = -\frac{\mu \beta^{}_2 g \omega}{\sqrt{\kappa}},
\end{equation}
where the dimensionless voltage fluctuation is given by 
\begin{equation}\label{eq:supp:g}
    g = \frac{1}{\sqrt{\left(\mu\frac{A}{g}\omega_{}^2\sin\theta-\beta^{}_2\omega\right)_{}^2+\left(\mu\frac{A}{g}\omega_{}^2\cos\theta-\beta^{}_1\right)_{}^2}}
\end{equation}
and we introduced the parameter $\kappa$
\begin{equation}\label{eq:supp:kappa}
    \kappa = \left(\frac{2\omega}{Q}\right)_{}^2+\left(1-(1+\mu_{}^2)\omega_{}^2\right)_{}^2 .
\end{equation}

\subsection{Current Function}

%
%
%

\begin{figure}[htb!]
\begin{centering}
    \includegraphics[width=70mm]{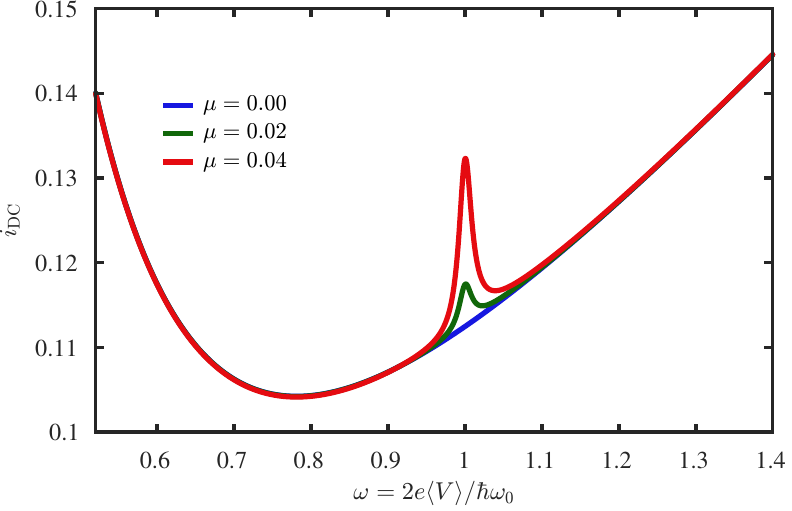}
    \caption{Plots of $i^{}_\text{DC}(\omega,\mu)$ (Eq.\,\ref{eq:supp:IVC}) for the parameters $\beta^{}_1=0.1,\ \beta^{}_2=2,\ Q=10^2_{}$ and coupling strengths $\mu=\{0.00,0.02,0.04\}$ (blue, green and red lines respectively). The low value of $Q$ is chosen to aid visibility.}
    \label{fig:idcplots}
\end{centering}
\end{figure}

The current function Eq.\,\ref{eq:supp:IVC} may be written in the form
\begin{equation}
    i^{}_\text{DC}(\omega,\mu) = i^{(0)}_\text{DC}(\omega) + \delta i^{}_\text{DC}(\omega,\mu),
\end{equation}
and is sketched in Fig.\ref{fig:idcplots} for different values of $\mu$. States with negative differential resistance $\partial i^{}_\text{DC}/\partial\omega<0$ are  unphysical (see Fig.2(b) in the main paper).

The uncoupled solution $i^{(0)}_\text{DC}(\omega)$ is found by evaluating the Eq.\,\ref{eq:supp:IVC} at $\mu=0$, yielding
\begin{equation}\label{eq:supp:IVC0}
	i_\text{DC}^{(0)}(\omega) = \beta^{}_1 \left(\omega + \frac{1}{2\omega(\beta_1^2+\beta_2^2 \omega_{}^2)}\right).
\end{equation}
This produces a ohmic behaviour ($i_\text{DC}^{(0)}(\omega)\rightarrow \beta^{}_1 \omega$) at large $\omega$, and has a global current minimum at $\omega \approx (3/2)^{1/4}/\sqrt{\beta_2}$ for $\beta_2\gg\beta_1$.

\begin{figure}[htb!]
\begin{centering}
    \includegraphics[width=70mm]{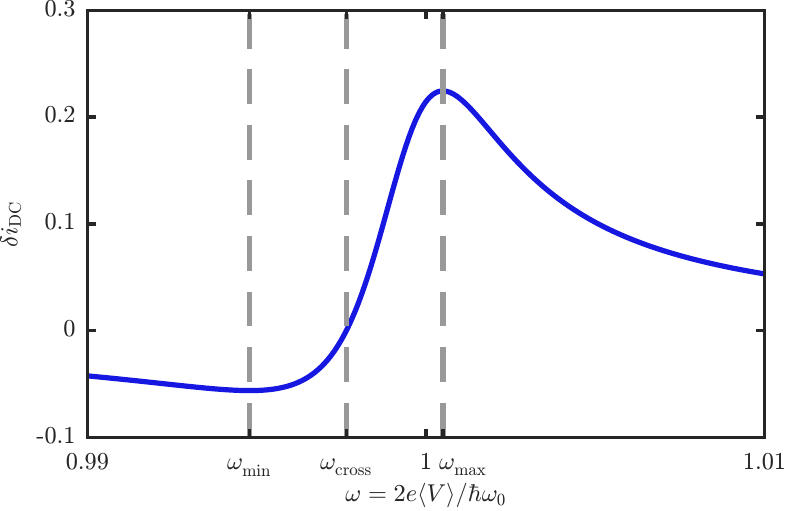}
    \caption{Plot of the expanded resonance $\delta i^{}_\text{DC}(\omega,\mu)$ (Eq.\,\ref{eq:supp:dIVC_Fano}, blue solid line) for parameters $\beta^{}_1=1,\ \beta^{}_2 = 2,\ Q = 10^3_{}$ and $\mu = 0.08$.
    The frequencies $\omega^{}_\text{min},\ \omega^{}_\text{cross}$ and $\omega^{}_\text{max}$
    are labelled and correspond to the grey dashed lines.}
    \label{fig:supp:didcplot}
\end{centering}
\end{figure}

The mechanically-induced correction $\delta i^{}_\text{DC}(\omega,\mu)$ can be written in the form
\begin{equation}\label{eq:supp:dIVC}
    \delta i^{}_\text{DC}(\omega,\mu) = \frac{N_1^{} N_2^{}}{D_1^{} D_2^{}},
\end{equation}
where the numerator and denominator terms are given by
\begin{equation}
    N_1^{} = \mu_{}^2 \beta_2^{} \omega_{}^3,
\end{equation}
\begin{equation}
    N_2^{} = \beta^{}_1\beta^{}_2(\omega_{}^2-1)+\omega_{}^2\beta^{}_2 \left(\frac{\beta^{}_2}{Q}+\frac{\beta^{}_1\mu_{}^2}{2}\right)-\frac{\beta_1^2}{Q},
\end{equation}
\begin{equation}
    D_1^{} = \beta_1^2+\beta_2^2\omega_{}^2,
\end{equation}
\begin{equation}
\begin{gathered}
    D_2^{} = (\omega_{}^2-1)\left[(\beta_1^2+\beta_2^2\omega_{}^2)(\omega_{}^2-1)+2\omega_{}^2\mu_{}^2\beta_1^2\right]\\
    +\omega_{}^4 \left(\beta^{}_1\mu_{}^2+\frac{2\beta^{}_2}{Q}\right)_{}^2+\beta_1^2 \left(\frac{2\omega}{Q}\right)_{}^2.
\end{gathered}
\end{equation}
These expressions are simplified by expanding in the vicinity of the mechanical resonance frequency ($\omega\simeq 1$) and assuming both weak coupling and large mechanical quality factor ($\mu\ll 1$ and $1/Q\ll 1$), resulting in the expansion
\begin{equation}\label{eq:supp:Expansion}
    \omega_{}^2 \rightarrow 1 + \epsilon \delta \omega_{}^2 \qq{,} \mu_{}^2 \rightarrow \epsilon \mu_{}^2 \qq{,} \frac{1}{Q} \rightarrow \frac{\epsilon}{Q}.
\end{equation}
Retaining the lowest order terms in $\epsilon$, the mechanical correction in Eq.\,\ref{eq:supp:dIVC} is  approximated by the Fano resonance
\begin{equation}\label{eq:supp:dIVC_Fano}
    \delta i^{}_\text{DC}(\omega,\mu) \approx \frac{\mu_{}^2 \beta^{}_1 \beta_2^2}{(\beta_1^2+\beta_2^2)_{}^2}\left[\frac{(\omega_{}^2-\omega_\text{Fano}^2)+\frac{q_{}^2-1}{q\Gamma}\left(\frac{\Gamma}{2}\right)_{}^2}{(\omega_{}^2-\omega_\text{Fano}^2)^2+\left(\frac{\Gamma}{2}\right)_{}^2}\right],
\end{equation}
where the Fano frequency, linewidth and Fano factor are given respectively by
\begin{equation}\label{eq:supp:fanofreq}
   \omega_\text{Fano}^2 = 1-\frac{\mu_{}^2}{1+q_{}^2},
\end{equation}
\begin{equation}\label{eq:supp:fanowidth}
   \frac{\Gamma}{2} = \frac{2}{Q}+\frac{\beta_2 \mu_{}^2}{1+q_{}^2},
\end{equation}
\begin{equation}
    q = \frac{\beta^{}_2}{\beta^{}_1}.
\end{equation}
The frequencies corresponding to the minimum, maximum and zero crossing of $\delta i^{}_\text{DC}(\omega,\mu)$ are given by
\begin{equation}\label{eq:supp:omegamin}
    \omega^{}_{\text{min}} = \sqrt{\omega_{\text{Fano}}^2-\frac{\Gamma q}{2}}
    = \sqrt{1-\mu_{}^2-\frac{2\beta^{}_2}{\beta^{}_1 Q}},
\end{equation}
\begin{equation}\label{eq:supp:omegamax}
    \omega^{}_{\text{max}} = \sqrt{\omega_{\text{Fano}}^2+\frac{\Gamma}{2q}} = 
    \sqrt{1+\frac{2\beta^{}_1}{\beta^{}_2 Q}},
\end{equation}
\begin{equation}\label{eq:supp:omegacross}
    \omega^{}_{\text{cross}} = \sqrt{\omega_{\text{Fano}}^2+\frac{\Gamma}{4}\left( \frac{1}{q}-q\right)} = \sqrt{1-\frac{\mu_{}^2}{2}+\frac{1}{Q}\left(\frac{\beta^{}_1}{\beta^{}_2}-
    \frac{\beta^{}_2}{\beta^{}_1}\right)},
\end{equation}
respectively, and are illustrated in Fig.\,\ref{fig:supp:didcplot}.
\\
In the limit $q\gg1$ (\textit{i.e.} $\beta^{}_2\gg\beta^{}_1$, which occurs with experimentally relevant parameters), the Fano-resonance Eq.\,\ref{eq:supp:dIVC_Fano} reduces to the Lorentzian expression 
\begin{equation}\label{eq:supp:didclorentzian}
    \delta i_{\text{DC}}(\omega,\mu)\approx\frac{\mu_{}^2}{\beta^{}_2\Gamma}\frac{1}{(\omega_{}^2-1)_{}^2 (\frac{2}{\Gamma})_{}^2+1} .
\end{equation}
In this regime it is simple to calculate the plateau width by evaluating $\delta i^{}_{\text{DC}}$ at its peak
\begin{equation}\label{eq:supp:didcmax}
    \Delta i^{}_{\text{DC}}(\mu)= \delta i^{}_{\text{DC}}(\omega^{}_{\text{max}},\mu) =  \frac{\beta_2^3}{(\beta_1^2+\beta_2^2)_{}^2}\frac{\mu_{}^2}{\frac{2\beta^{}_1\beta^{}_2\mu_{}^2}{\beta_1^2+\beta_2^2}+\frac{4}{Q}}.
\end{equation}
For $\mu\ll 2(1+q^{2}_{})/Qq$ this reduces to 
\begin{equation}\label{eq:supp:didcmaxsmallmu}
    \Delta i^{}_{\text{DC}}(\mu) \approx \frac{\mu_{}^2 Q \beta_2^3}{4(\beta_1^2+\beta_2^2)_{}^2},
\end{equation}
whereas for $\mu\gg 2(1+q^{2}_{})/Qq$ the plateau width saturates to the value $\Delta i^{}_{\text{DC}}=\beta^{2}_2/2\beta^{}_1(\beta_1^2+\beta_2^2)$. 
In reality however, at strong coupling plateau states become energy unstable and the plateau cannot be explored entirely.

\subsection{Resonator Amplitude}

\begin{figure}[htb!]
\begin{centering}
    \includegraphics[width=70mm]{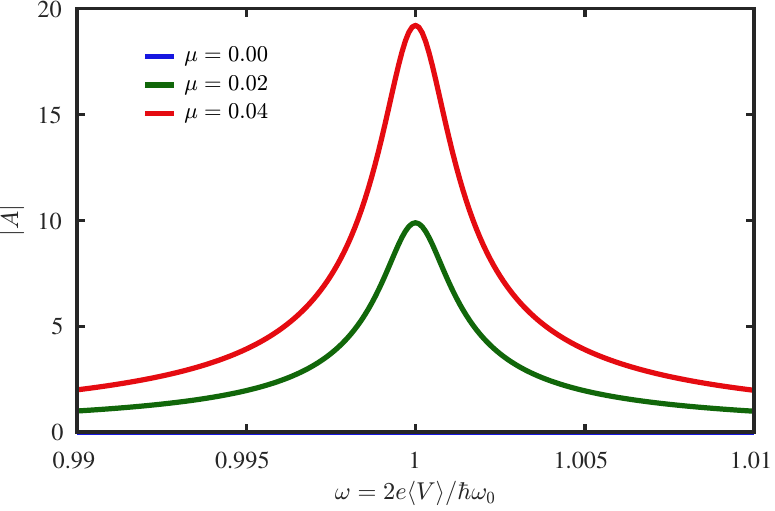}
    \caption{Plots of the mechanical amplitude $\abs{A}$ (Eq.\,\ref{eq:supp:A}) for the parameters $\beta^{}_1=0.1,\ \beta^{}_2=2,\ Q=10^3_{}$, and the coupling strengths $\mu=\{0.00,0.02,0.04\}$ (blue, green and red lines respectively).}
    \label{fig:supp:Aplots}
\end{centering}
\end{figure}

Performing the expansion (Eq.\,\ref{eq:supp:Expansion}) on $A(\omega,\mu)$ from Eq.\,\ref{eq:supp:A} reveals that the mechanical amplitude is approximately Lorentzian
\begin{equation}\label{eq:supp:Aapprox}
    A(\omega,\mu) \approx -\frac{\mu\beta^{}_2}{\sqrt{\beta_1^2+\beta_2^2}}\frac{1}{\sqrt{\left(\omega_{}^2-\omega_\text{Fano}^2\right)_{}^2+\left(\frac{\Gamma}{2}\right)_{}^2}}.
\end{equation}
The amplitude is therefore maximal at the Fano frequency, where it is equal to 
\begin{equation}\label{eq:supp:Amax}
    A^{}_\text{max}(\mu) = \abs{A(\omega^{}_\text{Fano},\mu)} = \frac{\abs{\mu}\beta^{}_2}{\sqrt{\beta_1^2+\beta_2^2}}\frac{1}{\frac{\beta^{}_1\beta^{}_2\mu_{}^2}{\beta_1^2+\beta_2^2}+\frac{2}{Q}}.
\end{equation}
Plots of $A(\omega,\mu)$ for various $\mu$ are shown in Fig.\,\ref{fig:supp:Aplots}.

\subsection{Validity of Approximations}

\begin{figure}[htb!]
\begin{centering}
    \includegraphics[width=70mm]{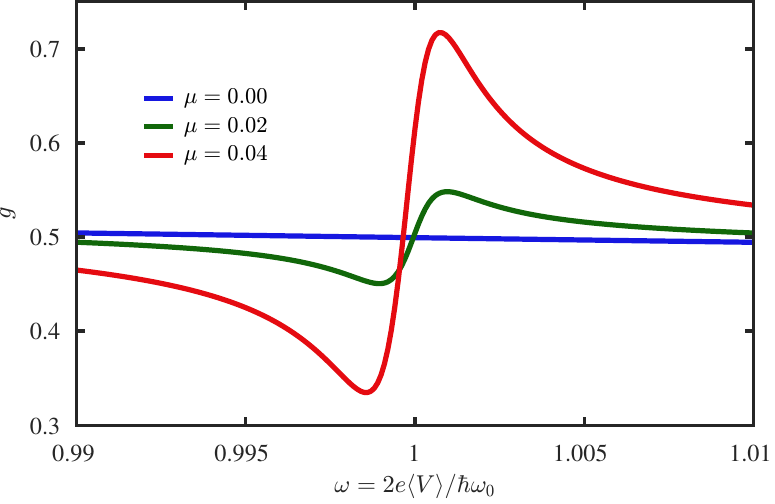}
    \caption{Plots of the dimensionless voltage fluctuation amplitude $g$ (Eq.\,\ref{eq:supp:g}) for the parameters $\beta^{}_1=0.1,\ \beta^{}_2=2,\ Q=10^3_{}$, and the coupling strengths $\mu=\{0.00,0.02,0.04\}$ (blue, green and red lines respectively).}
    \label{fig:supp:gplots}
\end{centering}
\end{figure}

The amplitude of voltage fluctuations $g$ is plotted in Fig.\,\ref{fig:supp:gplots} for multiple values of $\mu$, revealing a Fano-like form. 
Expanding $g$ via Eq.\,\ref{eq:supp:Expansion} yields
\begin{equation}\label{eq:supp:gapprox}
    g_{}^2 (\omega,\mu) \approx \frac{1}{\beta_1^2+\beta_2^2}\frac{[1-(1+\mu_{}^2)\omega_{}^2]_{}^2+\left(\frac{2\omega}{Q}\right)_{}^2}{(\omega_{}^2-\omega_{\text{Fano}}^2)_{}^2+\left(\frac{\Gamma}{2}\right)_{}^2}.
\end{equation}
Whilst the maximum of this expression occurs at $\omega>\omega^{}_\text{max}$, the largest physical solution occurs at $\omega^{}_\text{max}$ itself.
In the Lorentzian regime $q\rightarrow\infty$ the maximum frequency occurs at approximately unity, such that the maximum value of $g$ is equal to  
\begin{equation}\label{eq:supp:gmax}
   g^{}_{\text{max}}(\mu)=g(1,\mu)\approx \frac{1}{\beta^{}_2}\sqrt{1+\left(\frac{\mu_{}^2 Q}{2}\right)_{}^2}.
\end{equation}
The ratio $g/\omega$ can therefore become larger than unity for $\mu_{}^2 \gtrsim 2\sqrt{\beta_2^2-1}/Q$.
This violates the approximation $g/\omega \ll 1$ such that our analytical approach becomes quantitatively incorrect, while good qualitative agreement remains. 

\section{Critical Couplings}

\begin{figure}[htb!]
\begin{centering}
    \includegraphics[width=70mm]{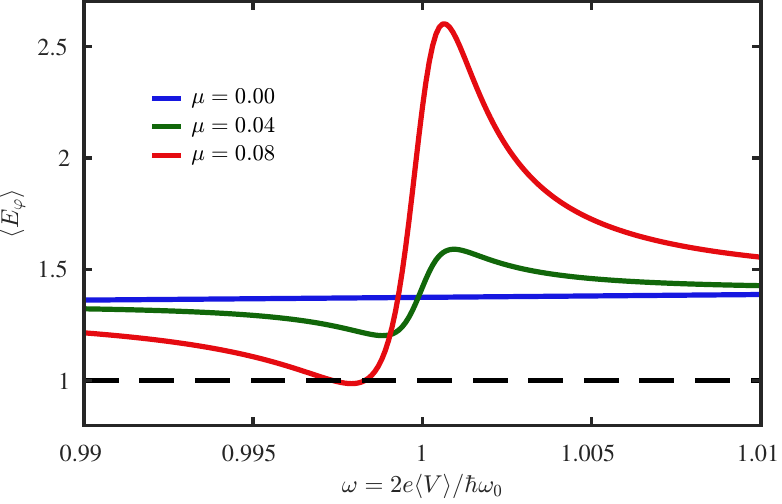}
    \caption{Plots of the dimensionless time-averaged electronic energy $\expval{E^{}_\varphi}$ (Eq.\,\ref{eq:supp:average_energy}) for the parameters $\beta^{}_1=0.1,\ \beta^{}_2=2,\ Q=10^3_{}$ and the coupling strengths $\mu=\{0.00,0.04,0.08\}$ (blue, green and red lines respectively).
    For $\mu>\mu^{}_\text{c1}$ (red line) the energy falls below the retrapping criteria $\expval{E^{}_\varphi}=1$ (black dashed line) causing mechanically-induced retrapping.}
    \label{fig:supp:energyplot}
\end{centering}
\end{figure}

The electronic energy of the system is equal to 
\begin{equation}\label{eq:supp:electronic_energy}
    E^{}_{\varphi} = \frac{\beta_2}{2}\dot{\varphi}^2-\cos\varphi.
\end{equation}
Inserting the ansatzes Eqs.\,\ref{eq:supp:phaseansatz} and \ref{eq:supp:dissplacementansatz} into this expression and taking the time average gives
\begin{equation}\label{eq:supp:average_energy}
    \expval{E_{\varphi}} = \frac{\beta^{}_2}{2}\left(\omega_{}^2+\frac{g_{}^2}{2}\right)-\frac{g}{2\omega}\sin{\varphi^{}_0}.
\end{equation}
Plotting Eq.\,\ref{eq:supp:average_energy} in Fig.\,\ref{fig:supp:energyplot} shows that finite $\mu$ causes a Fano-like resonance in $\expval{E_{\varphi}}$, the negative lobe of which is responsible for mechanically induced retrapping.

Fixing $\expval{E^{}_{\varphi}}=1$ defines the contour separating energy-stable and unstable states.
This contour is determined by the quadratic equation in $\mu^2$
\begin{equation}\label{eq:supp:ContourExact}
    a\mu_{}^4+b\mu_{}^2+c=0
\end{equation}
where
\begin{equation}
    \frac{a}{\omega_{}^4} = \beta_1^2\left(\omega_{}^2-\frac{2}{\beta^{}_2}\right)+\frac{1}{2},
\end{equation}
\begin{equation}
    \frac{b}{2\omega_{}^2} = \left(\omega_{}^2-\frac{2}{\beta^{}_2}\right)\left[\beta_1^2(\omega_{}^2-1)+\frac{2\beta^{}_1 \beta^{}_2}{Q}\omega_{}^2\right]+(\omega_{}^2-1),
\end{equation}
\begin{equation}
    c = \left[ \left( \omega_{}^2-\frac{2}{\beta^{}_2}\right)(\beta_1^2+\beta_2^2\omega_{}^2)+\frac{3}{2}\right] \left[ (\omega_{}^2-1)_{}^2+\left(\frac{2\omega}{Q}\right)_{}^2\right].
\end{equation}
The equation of the contour is simplified to
\begin{equation}\label{eq:supp:ExpandedContourShort}
    \Tilde{a}\mu^4 + \Tilde{b}\mu^2 + \Tilde{c}=0
\end{equation}
by expanding $a,\ b$ and $c$ to first order in $\epsilon$ using the expansion Eq.\,\ref{eq:supp:Expansion}
\begin{equation}
    \Tilde{a} = \beta_1^2\left(1-\frac{2}{\beta^{}_2}\right)+\frac{1}{2} = a^{}_0,
\end{equation}
\begin{equation}
    \Tilde{b} = 2\left(1-\frac{2}{\beta^{}_2}\right)\left[\beta_1^2(\omega_{}^2-1)+\frac{2\beta^{}_1 \beta^{}_2}{Q}\right]+2(\omega_{}^2-1) = b^{}_0+b^{}_1 (\omega_{}^2-1),
\end{equation}
\begin{equation}
    \Tilde{c} = \left[ \left( 1-\frac{2}{\beta^{}_2}\right)(\beta_1^2+\beta_2^2)+\frac{3}{2}\right] \left[ (\omega_{}^2-1)_{}^2+\left(\frac{2}{Q}\right)_{}^2\right] = c^{}_0+c^{}_2 (\omega_{}^2 -1)_{}^2 .
\end{equation}
This leads to the quadratic equation in $(\omega^2-1)$
\begin{equation}\label{eq:supp:ExpandedContourEqn}
    c^{}_2 (\omega_{}^2-1)_{}^2 +\mu_{}^2 b^{}_1 (\omega_{}^2-1)+a^{}_0 \mu_{}^4 +b^{}_0 \mu_{}^2 +c^{}_0 = 0.
\end{equation}

\subsection{First Critical Coupling}

The smallest value of $\mu$ at which real solutions of Eq.\,\ref{eq:supp:ExpandedContourEqn} exist is the  first critical coupling $\mu^{}_\text{c1}$.
An expression for $\mu^{}_\text{c1}$ is derived by setting the discriminant of Eq.\,\ref{eq:supp:ExpandedContourEqn} to zero
\begin{equation}\label{eq:supp:mu_c1_quadratic}
    p \mu_{\text{c}1}^2 +q \mu^{}_{\text{c}1} +r = 0,
\end{equation}
where 
\begin{equation}
    p = b_1^2-4a^{}_0 c^{}_2,
\end{equation}
\begin{equation}
    q = -4b^{}_0 c^{}_2,
\end{equation}
\begin{equation}
    r = -4 c^{}_0 c^{}_2.
\end{equation}
Solving Eq.\,\ref{eq:supp:mu_c1_quadratic} yields an expression for $\mu^{}_\text{c1}$
\begin{equation}\label{eq:supp:mu_c1_TwoRoots}
    \mu_{\text{c}1}^2 =  \frac{4}{Q\beta_2}\cdot \frac{(\beta^{}_2-2)(\beta_1^2+\beta_2^2)+\frac{3}{2}\beta^{}_2}{\sqrt{1-2\beta^{}_2(\beta^{}_2-2)}-2\beta^{}_1(\beta^{}_2-2)}.
\end{equation}
In the experimentally relevant limit $\beta^{}_2\gg \beta^{}_1$ this may be approximated by 
\begin{equation}\label{eq:supp:muc1}
    \mu^2_\text{c1}\approx\frac{4\beta^{}_2(\beta^{}_2-2)+6}{Q\sqrt{1-2\beta^{}_2(\beta^{}_2-2)}}.
\end{equation}
%
%
%
%
A formula for the maximum plateau width may be derived by substituting $\mu^{}_\text{c1}$ into the plateau width Eq.\,\ref{eq:supp:didcmaxsmallmu} to produce 
\begin{equation}
    \Delta i^\text{max}_\text{DC}= \delta i^{}_\text{DC}(\omega^{}_\text{max},\mu^{}_\text{c1})\simeq
    \frac{2\beta^{}_2(\beta^{}_2-2)+3}{2\beta^{}_2\sqrt{1-2\beta^{}_2(\beta^{}_2-2)}}.
\end{equation}

\subsection{Second Critical Coupling}

The second critical coupling $\mu^{}_\text{c2}$ is the value of $\mu$ at which the transition frequency $\omega^{}_2$ becomes energy-unstable, preventing shoulder access via the decreasing current bias path.
An expression for $\mu^{2}_\text{c2}$ is derived by finding the intersection of the energy-stability contour Eq.\,\ref{eq:supp:ExpandedContourEqn} with $\omega^{}_2$.
In the limit $Q\gg1$ this frequency is well approximated by $\omega^{}_2\approx\omega^{}_\text{cross}$, and substituting this value yields the equation
\begin{equation}
    \left[\frac{c^{}_2}{4}-\frac{1}{2}\right]\mu^{4}_\text{c2}+
    \left[b^{}_0+\frac{c^{}_2-b^{}_1}{Q}\left(\frac{\beta^{}_2}{\beta^{}_1}-\frac{\beta^{}_1}{\beta^{}_2}\right)\right]\mu^{2}_\text{c2}+
    \frac{c^{}_2}{Q}\left[4+\left(\frac{\beta^{}_2}{\beta^{}_1}-\frac{\beta^{}_1}{\beta^{}_2}\right)_{}^2\right]=0.
\end{equation}
In the regime $\beta^{}_2\gg\beta^{}_1$ this has the approximate solution
\begin{equation}\label{eq:supp:muc2}
    \mu^2_\text{c2}\approx \frac{2\beta^{}_2}{\beta^{}_1 Q}\left[\frac{2}{\sqrt{1-2\beta^{}_2(\beta^{}_2-2)}}-1\right].
\end{equation}

\subsection{Third Critical Coupling}

Above the third critical coupling $\mu^{}_\text{c3}$ there are no energy-stable solutions on the mechanical branch.
Quantitatively, $\mu^{}_\text{c3}$ is therefore solved via the intersection of the energy-unstable contour and the global current minimum where $\partial i^{}_\text{DC}/\partial\omega=0$.
For large $Q$ and small $\mu$ this frequency is negligibly renormalised by the mechanical coupling, and is approximated by the minimum of $i^{(0)}_\text{DC}$, which for $\beta^{}_2\gg\beta^{}_1$ is equal to $\omega\approx(3/2)^{1/4}/\sqrt{\beta^{}_2}$.
As this frequency is not necessarily in the resonant region, and may therefore require larger $\mu$ to make this frequency unstable, neither $(\omega_{}^2-1)$ nor $\mu_{}^2$ can be regarded as small parameters, and the expansion of the contour via Eq.\,\ref{eq:supp:Expansion} is not valid.
Instead, the exact contour equation Eq.\,\ref{eq:supp:ContourExact} is taken and the small parameters $1/Q$ and $\beta^{}_1$ are neglected before the minimum frequency is substituted to deduce an equation for $\mu^{}_\text{c3}$
\begin{equation}
    \mu^4_\text{c3}+4\left(1-\beta^{}_2\sqrt{\frac{2}{3}}\right)\mu^2_\text{c3}+
    6\left(1-\sqrt{\frac{2}{3}}\right)\left(1-\beta^{}_2\sqrt{\frac{2}{3}}\right)_{}^2=0.
\end{equation}
Although this equation has two solutions, one is unphysically large and the lower solution is chosen
\begin{equation}
    \mu^{2}_\text{c3}\approx\sqrt{2/3}\left(2-\sqrt{2(\sqrt{6}-1)}\right)\left(\beta^{}_2-\sqrt{3/2}\right).
\end{equation}
The numerical prefactor may be approximated as $0.493\approx 1/2$ to produce the final expression 
\begin{equation}\label{eq:supp:muc3}
    \mu^{}_\text{c3}\approx \frac{1}{2}\sqrt{\beta^{}_2-\sqrt{3/2}}.
\end{equation}

\subsection{Comparison With Simulations}

\begin{figure}[htb!]
\begin{centering}
    \includegraphics[width=70mm]{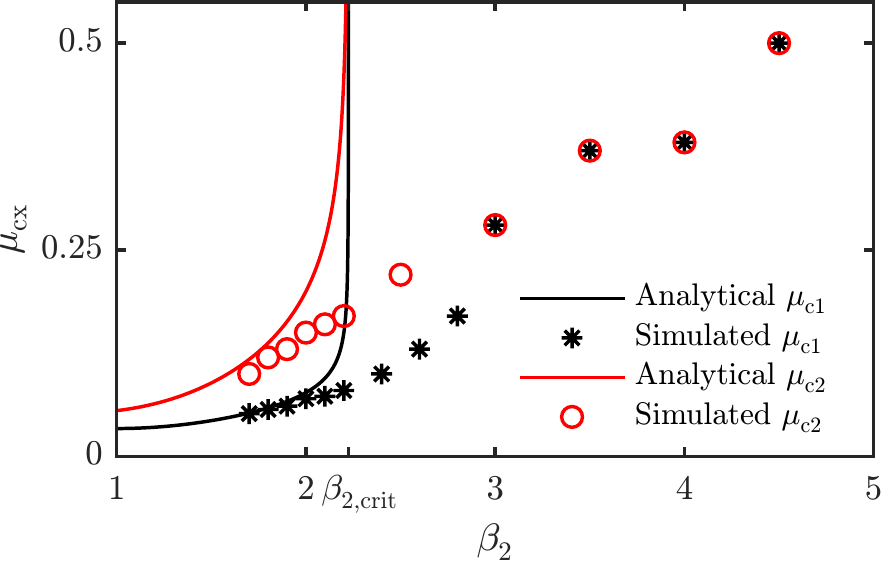}
    \caption{Comparison of simulated values of $\mu^{}_\text{c1}$ (black asterisks) and $\mu^{}_\text{c2}$ (red circles) with the derived analytical expressions Eqs.\,\ref{eq:supp:muc1} and \ref{eq:supp:muc2} (black and red solid lines respectively), for $\beta^{}_1=0.1$, $Q=10^3_{}$ and a range of $\beta^{}_2$.
    }
    \label{fig:supp:mu_c12_compare}
\end{centering}
\end{figure}

The analytical expressions for $\mu^{}_\text{c1}$ and $\mu^{}_\text{c2}$ (Eqs.\,\ref{eq:supp:muc1} and \ref{eq:supp:muc2}) are compared with numerically simulated values as a function of $\beta^{}_2$ in Fig.\,\ref{fig:supp:mu_c12_compare}, showing good agreement for small $\beta^{}_2$.
This agreement breaks down for $\beta^{}_2\gtrsim\beta^{}_\text{2,crit}$. The analytical estimate $\beta^{}_\text{2,crit}\approx 1+\sqrt{3/2}$, stemming from the assumptions $\mu\ll 1$ and $g/\omega\ll 1$, is lower than the numerical one $\beta^\text{num}_\text{2,crit}\approx 5$.
Note also that the numerical $\mu^{}_\text{c1}$ appears to diverge sooner than $\mu^{}_\text{c2}$, but the former is capped to the latter as shoulder entry is necessary for retrapping on the increasing $i^{}_\text{DC}$ path.

\begin{figure}[htb!]
\begin{centering}
    \includegraphics[width=70mm]{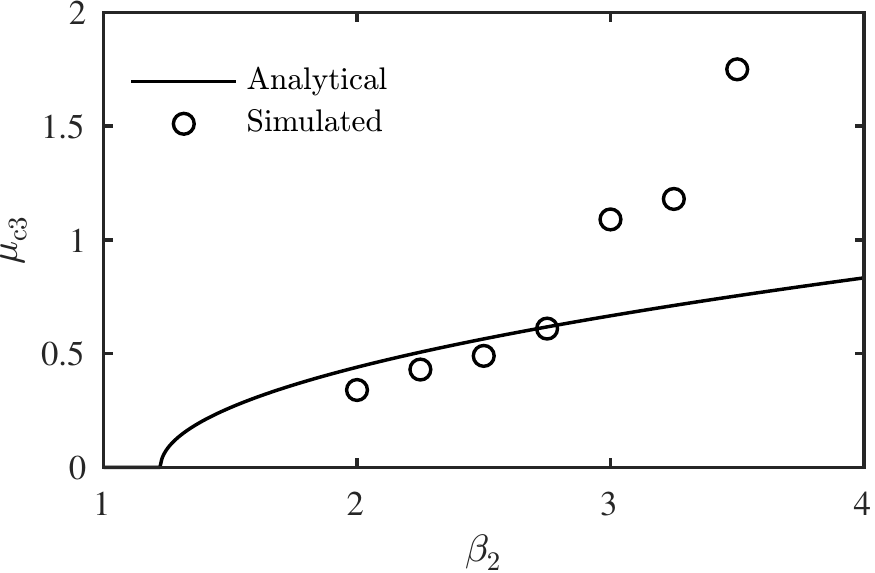}
    \caption{Comparison of simulated values of $\mu^{}_\text{c3}$ (circles) with the analytical expression Eq.\,\ref{eq:supp:muc3} (solid line) for $\beta^{}_1=0.1$, $Q=10^3_{}$ and a range of values of $\beta^{}_2$.}
    \label{fig:supp:mu_c3_compare}
\end{centering}
\end{figure}

A similar comparison of the analytical expression for $\mu^{}_\text{c3}$ Eq.\,\ref{eq:supp:muc3} and numerically simulated values is shown in Fig.\,\ref{fig:supp:mu_c3_compare}.
Whilst the agreement is good for moderate values of $\beta^{}_2$ this degrades at large values, as a result of neglecting the effect of $\mu$ on the global current minimum.
%

\begin{figure}[htb!]
\begin{centering}
    \includegraphics[width=70mm]{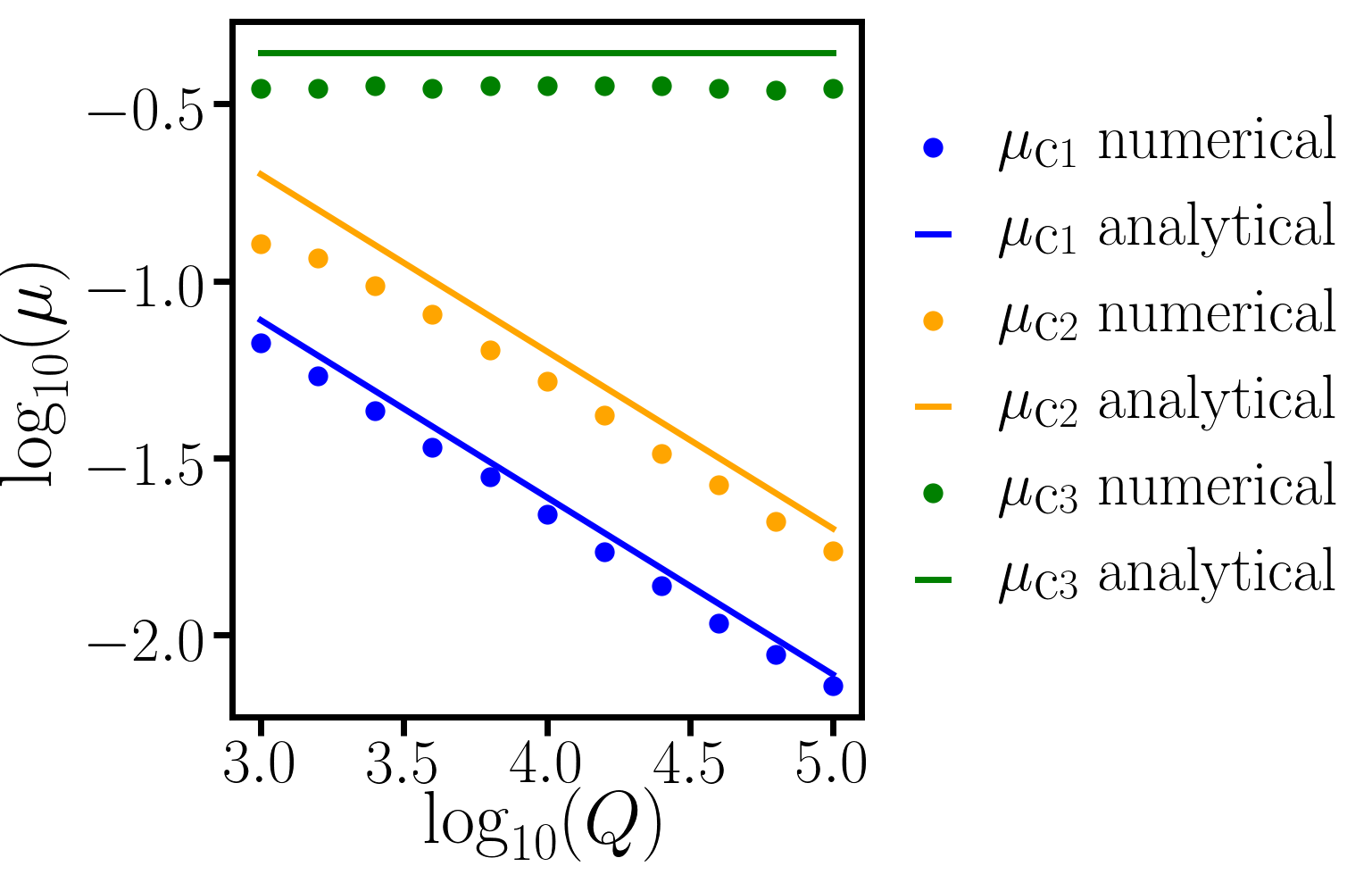}
    \caption{Comparison of simulated values (dots) and analytical formulas for $\mu^{}_\text{c1}$ (Eq.\,\ref{eq:supp:muc1}, blue), $\mu^{}_\text{c2}$ (Eq.\,\ref{eq:supp:muc2}, yellow) and $\mu^{}_\text{c3}$ (Eq.\,\ref{eq:supp:muc3}, green) for $\beta^{}_1=0.1,\ \beta^{}_2=2$ and a range of $Q$ (double log-scale).}
    \label{fig:supp:mubarQ}
\end{centering}
\end{figure}

Finally, the analytical expressions for all three critical couplings are compared with numerically simulated values as a function of $Q$ in Fig.\,\ref{fig:supp:mubarQ}.
The result is plotted on a double log-scale, confirming that the scaling with $Q$ predicted by the analytical expressions is correct.

\section{Experimental and Numerical Procedures}

\subsection{Experimental Procedure}

 In order to extract the parameters (resonance frequency and quality factor) of the suspended resonator from the experimental IVC, one should first determine the critical current $I^{}_\text{c}$, electrical resistance $R$ and Stewart-McCumber parameter $\beta^{}_\text{c}$ from the purely uncoupled measurements ($\mu=0$).
Turning on the mechanical coupling, the mechanically-induced plateau voltage may be measured to characterise the mechanical resonance frequency $\omega^{}_0\approx 2eV^{}_0/\hbar$ (accurate to a very high degree for underdamped suspended resonators with large $Q$).
The parameters $\beta^{}_1=\omega^{}_0/\omega^{}_\text{c}$ and $\beta^{}_2=\beta^2_1 \beta^{}_\text{c}$ may then be deduced. 

To determine the magnetic field scale $B^{}_0$ from Eq.\,\ref{eq:supp:muc3}, the increasing magnetic field procedure must be performed to determine the third critical magnetic field $B^{}_\text{c3}$.
Finally, the procedure of increasing $i^{}_\text{DC}$ from the mechanically-induced shoulder must be performed to determine the first critical coupling $\mu^{}_\text{c1}=B^{}_\text{c1}/B^{}_0$, leading to the value of $Q$ via Eq.\,\ref{eq:supp:muc1}.

As the experimental procedure consists only of transport measurements, the mechanical amplitude of the suspended junction must be computed analytically.
We cannot use the expression $A(\omega,\mu)$ in  Eq.\,\ref{eq:supp:A} however as this assumes voltage bias; for large $Q$ the plateau voltage is quasi-constant, and it would be experimentally challenging to measure the small changes in voltage necessary to calculate $A$ accurately.
In the experimentally relevant limit $\beta^{}_2\gg \beta^{}_1$ however, Eqs.\,\ref{eq:supp:didclorentzian} and \ref{eq:supp:Aapprox} can be combined to give
\begin{equation}
    A=2\sqrt{\frac{\beta^{}_2 \Delta i^{}_\text{DC}}{\Gamma}}
\end{equation}
where $\Delta i^{}_\text{DC}$ is the position along the plateau (measured from its root) in units of $I^{}_\text{c}$.
To make this amplitude dimensional it must be multiplied by the characteristic length scale $x^{}_0=B^{}_0 L I^{}_\text{c}/\omega^2_0$ (assuming that the resonator length $\ell$ has been characterised during fabrication).

\subsection{Numerical Procedure}

Deterministic simulations begin at zero current bias and magnetic field, with initial conditions $\varphi = \dot{\varphi} = a = \dot{a}=0$, before numerically solving Eqs.\,\ref{eq:supp:rcsj} and \ref{eq:supp:mech} (using a Runge-Kutta-Fehlberg solver) until they converge to a steady state solution.
As the electrical quality factor $1/\beta^{}_1$ is much less than the mechanical quality factor $Q$, the time-scale for transients to dissipate is dictated by $Q$, and a dimensionless run time  $\tau_{\text{run}}=2\times10_{}^4>Q$ (corresponding to a dimensional time of $20\si{\micro\second}$) is chosen.
%
%
After this time the dimensionless DC voltage  $\omega=2e\expval{V}/\hbar\omega^{}_0=\expval{\dot{\varphi}}$ and mechanical RMS $a^{}_\text{RMS} = \sqrt{\expval{a_{}^2}-\expval{a}_{}^2}$ are computed.
The parameters $i^{}_{\text{DC}}$ or $\mu$ are then incrementally changed by $10^{-4}_{}$ or $10^{-3}_{}$, respectively, and the procedure is repeated, using the values of $\{\varphi,\dot{\varphi},a,\dot{a}\}$ at $\tau^{}_\text{run}$ as initial conditions for the next iteration, until the IVC is generated.

Stochastic simulations are performed in a similar manner, but are solved using a explicit order 1.5 strong scheme with a fixed time-step $\Delta\tau=10^{-3}_{}$.